\documentclass{aa}  
\usepackage{orcidlink}
\usepackage{graphicx}
\usepackage{amsmath,siunitx}
\usepackage{natbib}
\bibpunct{(}{)}{;}{a}{}{,} % to follow the A&A style
\usepackage{color}
\usepackage{float}
\usepackage{hyperref}
\usepackage{longtable}
\usepackage{rotating}
\usepackage{lscape}
\usepackage{soul}
\usepackage{subcaption}
\usepackage{comment}

\usepackage{xcolor}
\usepackage{txfonts}

\hypersetup{
  colorlinks,
  citecolor=teal,
  linkcolor=cyan,
  urlcolor=violet}

\newcommand{\tuner}{\textsc{TUNER}}

\newcommand\Htwo{H$_{\rm 2}$}

\usepackage{textgreek}

\begin{document}

    \title{Hiding behind a curtain of dust: Gas and dust properties of an ultra-luminous strongly-lensed $z=3.75$ galaxy behind the Milky Way disc}

    \author{Belén Alcalde Pampliega\orcidlink{0000-0002-41400428}\inst{\ref{inst:ESO},\ref{inst:UDP},\ref{inst:SKAO}}\thanks{e-mail: Belen.AlcaldePampliega@skao.int}
    \and Kevin C. Harrington\orcidlink{0000-0001-5429-5762}\inst{\ref{inst:JAO},\ref{inst:NAOJ},\ref{inst:ESO},\ref{inst:UDP}}
    \and Aristeidis Amvrosiadis\orcidlink{0000-0002-2985-7994}\inst{\ref{inst:UDurham}}
    \and Manuel Aravena\orcidlink{0000-0002-6290-3198}\inst{\ref{inst:UDP},\ref{inst:MINGAL}}
    \and Min S. Yun\orcidlink{0000-0001-7095-7543}\inst{\ref{inst:UMass}}
    \and Hugo Messias\orcidlink{0000-0002-2985-7994}\inst{\ref{inst:ESO},\ref{inst:JAO}}
    \and Antonio Hernán-Caballero \orcidlink{0000-0002-4237-5500}\inst{\ref{inst:CEFCA}}
    \and Leindert Boogaard\orcidlink{0000-0002-3952-8588}\inst{\ref{inst:Leiden}}
    \and Axel Weiß \orcidlink{0000-0000-0000-0000}\inst{\ref{inst:MaxPlank}}
    \and Benjamin Beauchesne\orcidlink{000-0002-0443-6018}\inst{\ref{inst:UDurhamExtragal}, \ref{inst:UDurhamCosmo} }
    \and Alejandro Santamaría-Miranda\orcidlink{0000-0001-6267-2820}\inst{\ref{inst:UChile}}
    \and Monica Ivette Rodriguez\orcidlink{0000-0000-0000-0000}\inst{\ref{inst:IRAM}}
    \and Eric Jiménez-Andrade\orcidlink{0000-0002-2640-5917}\inst{\ref{inst:Mex}}
    \and Manuel Solimano\orcidlink{0000-0001-6629-0379}\inst{\ref{inst:UDP}}
    \and James Lowenthal\orcidlink{0000-0001-9969-3115}\inst{\ref{inst:CSmith}}
    \and Patrick Kamieneski \orcidlink{0000-0001-9394-6732}\inst{\ref{inst:AZTempe}}
    \and Q. Daniel Wang \orcidlink{0009-0002-9714-1928}\inst{\ref{inst:UMass}}
    \and Chentao Yang\orcidlink{0000-0002-8117-9991}\inst{\ref{inst:Chalmers}}
    \and Amit Vishwas\orcidlink{0000-0002-4444-8929}\inst{\ref{inst:UCornell}}
    \and Pascale Hibon \orcidlink{0000-0000-0000-0000}\inst{\ref{inst:ESO}}    
    \and Brenda Frye\orcidlink{0000-0003-1625-8009}\inst{\ref{inst:AZTucson}}
    \and Jorge González-Lopez\orcidlink{0000-0000-0000-0000}\inst{\ref{inst:UDP}}
    \and Yiqing Song\orcidlink{0000-0002-3139-3041}\inst{\ref{inst:ESO}}
    \and Meghana Killi\orcidlink{0000-0001-5289-3291}\inst{\ref{inst:UDP}} 
    }
    
    \institute{
    ESO Vitacura, Alonso de Córdova 3107,Vitacura, Casilla 19001, Santiago de Chile, Chile\label{inst:ESO}
    \and
    Instituto de Estudios Astrofísicos, Facultad de Ingeniería y Ciencias, Universidad Diego Portales, Av. Ejército Libertador 441, Santiago, Chile \label{inst:UDP}
    \and
    SKA Observatory, Jodrell Bank, SK11 9FT, UK\label{inst:SKAO}
    \and
    Joint ALMA Observatory, Alonso de Córdova 3107, Vitacura 763-0355, Santiago, Chile\label{inst:JAO}
    \and
    National Astronomical Observatory of Japan, Los Abedules 3085 Oficina 701, Vitacura 763 0414, Santiago, Chile\label{inst:NAOJ}
    \and
    Centre for Extragalactic Astronomy, Durham University, Department of Physics, South Road, Durham DH1 3LE, 
    UK \label{inst:UDurham}
    \and
    Millenium Nucleus for Galaxies (MINGAL)\label{inst:MINGAL}
    \and
    Department of Astronomy, University of Massachusetts, Amherst, MA 01003, USA \label{inst:UMass}
    \and
    Centro de Estudios de Física del Cosmos de Aragón (CEFCA), Plaza San Juan 1, E-44001 Teruel, Spain \label{inst:CEFCA}
    \and
    Leiden Observatory, Leiden University, P.O. Box 9513, 2300 RA Leiden, The Netherlands \label{inst:Leiden}
    \and
    Max-Planck-Institut für Radioastronomie, Auf dem Hügel 69, 53121 Bonn, Germany \label{inst:MaxPlank} 
    \and 
    Centre for Extragalactic Astronomy, Department of Physics, Durham University, South Road, Durham DH1 3LE, UK \label{inst:UDurhamExtragal}
    \and
    Institute for Computational Cosmology, Durham University, South Road, Durham DH1 3LE, UK \label{inst:UDurhamCosmo}
    \and
    Departamento de Astronomía, Universidad de Chile, Camino El Observatorio 1515, Las Condes, Santiago, Chile \label{inst:UChile}
    \and
    Institut de Radioastonomie Millimétrique (IRAM), Av. Divina Pastora 7, Núcleo Central 18012, Granada, Spain \label{inst:IRAM}
    \and
    Instituto de Radioastronomía y Astrofísica, Universidad Nacional Autónoma de México, Antigua Carretera a Pátzcuaro 8701, Ex-Hda. San José de la Huerta, Morelia, Michoacán, C.P. 58089, México\label{inst:Mex}
    \and
    Department of Astronomy, Smith College, Northampton, MA 01063, USA \label{inst:CSmith}
    \and
    School of Earth and Space Exploration, Arizona State University, PO Box 871404, Tempe, AZ 85287-1404, USA\label{inst:AZTempe}
    \and
    Department of Space, Earth \& Environment, Chalmers University of Technology, SE-412 96 Gothenburg, Sweden \label{inst:Chalmers}    
    \and
    Cornell Centre for Astrophysics and Planetary Science, Cornell University, Space Sciences Building, Ithaca, NY 14853, USA\label{inst:UCornell}
    \and
    Steward Observatory, University of Arizona, 933 N Cherry Ave, Tucson, AZ, 85721-0009 \label{inst:AZTucson}
    }

   \date{Received June 25, 2025; accepted December 05, 2025}

  \abstract{
  We present a detailed analysis of J154506, a strongly lensed submillimetre galaxy (SMG) behind the Lupus-I molecular cloud, 
  and a characterisation of its physical properties  
  using a combination of new and archival data, including VLT/MUSE and FORS2 optical data.
  We identify two high-significance (S/N >5) emission lines at 97.0 and 145.5 GHz, corresponding to CO(4-3) and CO(6-5), respectively, in spectral scans from the Atacama Compact Array (ACA) and the Large Millimetre Telescope (LMT), as well as the [CII] 158~$\mu$m fine-structure line at 400~GHz observed with the Atacama Pathfinder Experiment (APEX).
  These detections yield a spectroscopic redshift of $z_{\rm{spec}}=3.7515\pm 0.0005$.
  We also report the detection of [CI], HCN(4-3), and two H$_2\rm{O}^+$ transitions, further confirming the redshift and providing insights into the physical properties of  J154506.
  By modelling sub-arcsecond resolution ($\ang{;;0.75}$) ALMA Band 6 and 7 continuum data in the uv-plane, we derive an average magnification factor of $6.0 \pm 0.4$, and 
  our analysis reveals a relatively cold dust (38~K) in a starburst galaxy ($\sim900~\rm{M}_{\odot}yr^{-1}$) with a high intrinsic dust mass ($\sim2.5\times10^{9}~\rm{M}_{\odot}$) and infrared (IR) luminosity ($\sim6\times10^{12}~\rm{L}_{\odot}$).
  Non-local thermodynamic equilibrium radiative transfer modelling of the joint dust spectral energy distribution (SED) and CO line excitation suggests the dust continuum emission is primarily associated with relatively diffuse regions with molecular gas densities of $10^2-10^4\rm{cm}^{-3}$, rather than compact, high-pressure environments typical of extreme starbursts or active galactic nuclei (AGNs). 
  This interpretation is supported by the close-to-unity ratio between the dust and gas kinetic temperatures, which argues against highly energetic heating mechanisms.
  The CO excitation ladder peaks close to CO(5-4) and is dominated by slightly denser molecular gas.
  Our results underscore the unique power of far-IR and submillimetre observations to both uncover and characterise scarce, strongly lensed, high-redshift galaxies, even when they are obscured by foreground molecular clouds.
  }

   \keywords{galaxies:starburst --
                galaxies: high-redshift --
                submillimetre: galaxies --
                gravitational lensing: strong
               }

    \titlerunning {Hidden below the dust: A $z=3.75$ extremely bright strongly-lensed galaxy behind the MW clouds}
    \authorrunning{B. Alcalde Pampliega et. al}
    \maketitle

%%%%%%%%%%%%%%%%%%%%%%%%%%%%%%%%%%%%%%%%%%%%%%%%%%%
\section{Introduction}
%%%%%%%%%%%%%%%%%%%%%%%%%%%%%%%%%%%%%%%%%%%%%%%%%%%
Over the past two decades, (sub)millimetre surveys have revolutionised our understanding of galaxy formation and evolution by revealing an unexpected population of high-redshift, dust-obscured massive galaxies with intense star formation rates (SFR), the so-called submillimetre (submm) galaxies \citep[SMGs; see, e.g.][for a review]{Casey2014}. 
Extremely bright SMGs \citep[i.e. S$_{500\mu\rm{m}}>100$~mJy;][]{Negrello2017} provide a unique opportunity to study the 
interstellar medium (ISM) of galaxies due to their high luminosity, which is often enhanced by gravitational lensing.
This natural magnification enables detailed studies of star formation (SF) activity, dust, and molecular gas properties and dynamics, providing insights into the conditions that prevail in the early stages of galaxy evolution at scales and sensitivities otherwise unattainable at such distances. 
However, 
for high-redshift galaxy surveys in the so-called zone of avoidance (ZOA), 
this progress is hampered by Galactic absorption and contamination from Galactic brown dwarfs 
\citep[e.g.][]{Kraan-Korteweg2000,Woudt2004,Amores2012,Duplancic2024}. 
The ZOA covers $\sim$25\% of the distribution of optically visible galaxies 
and is reduced to 10-20\% in IR surveys \citep{Kraan-Korteweg2000,Kraan-Korteweg2005}. 
Because of their very red colours and point-like appearance, brown dwarfs mimic the observational properties of distant SMGs, complicating the identification and selection of genuine high-redshift sources. Furthermore, extremely bright SMGs and low-mass starless cores exhibit remarkably similar flux densities across the mid-infrared (MIR)-to-submm regime, adding another layer of complexity to disentangling these populations \citep[e.g.][]{Barnard2004, Wilkins2014}.
Studies of both local and high-redshift sources must therefore
account for each other as potential sources of contamination. In fact, the unprecedented sensitivity of the James Webb Space Telescope (JWST) has led to the identification of Galactic brown dwarfs in deep, multiband imaging and spectroscopic extragalactic surveys \citep[e.g.][]{Nonino2023, Hainline2024, Burgasser2024}.

Dust emission in a typical local galaxy peaks at wavelengths around $100\,\mu$m rest-frame.
At $z=3-4$, this peak will move into the 500\,$\mu$m SPIRE (Spectral and Photometric Imaging Receiver, on board the Herschel Space Observatory) band \citep[e.g. ][]{Dowell2014, Ivison2016, Clements2024}.
Sources whose dust emission has not reached its peak in this band
will most likely lie above redshift 4 \citep[e.g.][]{Greenslade2019, Greenslade2020}.
Understanding the intrinsic physical properties of strongly lensed galaxies requires precise knowledge of the redshifts of both the lens and the background lensed galaxy. 
Unfortunately, the optical-near-infrared (NIR) spectroscopic confirmation of dusty red galaxies at high-redshift, often extremely red or invisible at those wavelengths, is very challenging \citep[e.g. ][]{AlcaldePampliega2019, Wang2019, Williams2019}. This becomes increasingly difficult when the entire system is further obscured by local dust clouds within our own galaxy, as is the case for the high-redshift galaxy J154506.3-344317.9 (hereafter J154506 for brevity).
As a consequence, submm spectral-scan observations targeting bright CO and [CI] emission lines,  
which are not affected by dust extinction and can be directly associated with the background source,  represent a more efficient and widely used method that 
has been proven to be very successful in getting robust and unambiguous redshifts 
\citep[e.g.][]{Vieira2013, Strandet2016, Reuter2020, Neri2020, Urquhart2022, Chen2022}.

Recently, sub-arcsec resolution 
observations with the Atacama Large Millimeter/submillimeter Array
(ALMA) of pre-brown dwarf candidates in the Lupus-I molecular cloud uncovered optical-and-NIR undetected objects exhibiting far infrared (FIR) spectral energy distributions (SEDs) compatible with both young (pre)stellar objects and extragalactic sources \citep{SantamariaMiranda2021}. Among them, J154506, located at a 12” distance from the target source, stands out due to its extremely bright submm flux with a SED rising up to 500$\mu$m, and compelling evidence of being a strongly gravitationally lensed SMG seen through the Milky Way disc \citep{SantamariaMiranda2021}. 

In this work, 
we combine archival data with submm spectral scans to report the 
spectroscopic redshift confirmation of J154506
and provide an initial investigation of its dust and molecular gas properties. 
Throughout the paper, we adopt a flat Lambda cold dark matter ($\Lambda$CDM) cosmology with H$_{0}=$70~km s$^{-1}$ Mpc$^{-1}$, $\Omega_{\Lambda}=$0.7, $\Omega_{M}=$0.3, and a \cite{Chabrier2003} initial mass function. All the magnitudes refer to the AB
system \citep{Oke1983}.

\renewcommand{\arraystretch}{1.3}

%%%%%%%%%%%%%%%%%%%%%%%%%%%%%%%%%%%%%%%%%%%%%%%%%%%
\section{J154506: A strongly lensed system}
%%%%%%%%%%%%%%%%%%%%%%%%%%%%%%%%%%%%%%%%%%%%%%%%%%%

\begin{figure}
    \includegraphics[width=0.5\textwidth]{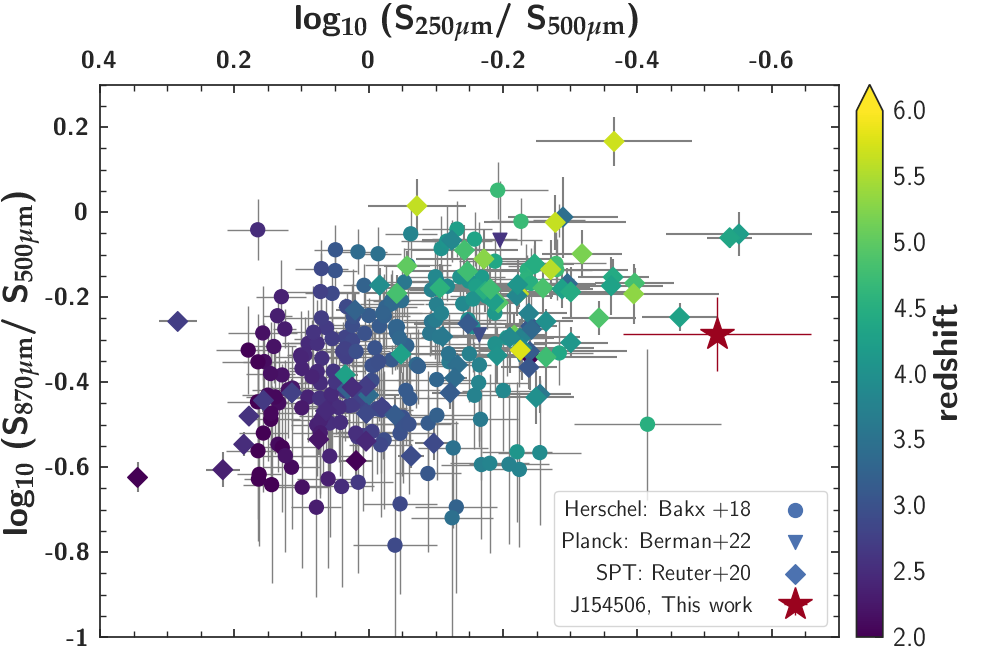}
   \caption{Far-infrared and submm colour-colour diagram (S$_{870\mu\rm{m}}$/S$_{500\mu\rm{m}}$ vs S$_{250\mu\rm{m}}$/S$_{500\mu\rm{m}}$), colour-coded by redshift. 
   The brightest known sources from the \textit{Herschel}, \textit{Planck}, and \textit{SPT} surveys are shown with circular, triangular, and rhomboidal symbols, respectively. 
   J154506 is highlighted with a red star. Flux densities at 850~$\mu$m (rather than at 870~$\mu$m) are used for the {Herschel} and \textit{Planck} sources. 
  The extremely red S$_{250\mu\rm{m}}$/S$_{500\mu\rm{m}}$ colour of  J154506 suggests a redshift of $z>3$.
   }
    \label{Fig:FIRcolours}
\end{figure}

J154506 (RA$=$15:45:06.333, DEC$=$-34:43:17.972) is a unique system 
located towards the Lupus-I Galactic molecular cloud in the Milky Way 
\citep[at a distance of $\sim153$~pc; e.g.][]{SantamariaMiranda2021}. It stands out owing to its extraordinarily bright (sub)mm flux \citep[S$_{500\mu \rm{m}}=134.4\pm$11.9 mJy; ][]{Tamura2015} and its extremely red S$_{250\mu\rm{m}}$/S$_{500\mu\rm{m}}$ colour (Fig.~\ref{Fig:FIRcolours}).
\cite{Tamura2015} reported that, based on the available multiwavelength observations (unresolved at that time), J154506 was likely not a star-like source, but instead a dusty galaxy at a cosmological distance. 

The far-infrared (FIR) and submm colours correlate with redshift, both empirically and theoretically \citep[e.g.][]{Burgarella2023, Cox2023}.
Fig.~\ref{Fig:FIRcolours} shows the S$_{870\mu\rm{m}}$/S$_{500\mu\rm{m}}$ versus S$_{250\mu\rm{m}}$/S$_{500\mu\rm{m}}$ colour-colour diagram
for some of the brightest known galaxies \citep[i.e. \textit{Herschel}, \textit{Planck}, and \textit{SPT} selected sources; ][]{Harrington2016,Reuter2020,Berman2022}, 
colour-coded by their spectroscopic redshift. 
There is a clear trend indicating that the brighter the source at 500~$\mu$m relative to 250~$\mu$m, the higher its redshift tends to be. 
According to its position in the diagram, J154506, highlighted with a red star, is also very likely located at $z>$3.
Given that the rest-frame FIR SED typically peaks around 100$\mu$m, 
the increase in flux density across the \textit{Herschel} SPIRE bands for J154506 
provides strong evidence for its classification as a high-redshift galaxy
with extreme IR luminosity. 
We also note that, unlike J154506, some of the highest redshift galaxy spectral energy distributions (SEDs) continue to rise or peak nearer the observed-frame 870$~\mu$m.
The limitations of submm colours and photometric redshifts, which provide only broad redshift ranges (as illustrated in Fig.~\ref{Fig:FIRcolours}), 
imply that precise and reliable redshift measurements require molecular and/or atomic FIR emission lines.

%-------------------------------------- 
%-------

Initial searches for Galactic molecular lines from J154506 had limited success; 
any detected $^{13}$CO(1-0) appeared homogeneous, and attempts to detect redshifted $^{12}$CO(2–1) failed to confirm its nature \citep{Tamura2015}.
While this manuscript was in preparation, \cite{Tamura2025} reported a redshift of $z = 3.753 \pm 0.001$ for J154506.  
Their work, based on observations with the Australian Telescope Compact Array, the Nobeyama 45m telescope, 
and ALMA, identified CO(2-1), CO(4-3), and CO(9-8) emission lines. 
Our observations independently detected CO(4-3), CO(6-5), [CI], [CII], and other tentative lines. 
This rich dataset enables a far more detailed and rigorous analysis of the gas excitation through
comprehensive combined modelling of dust and CO line SEDs. 
Our work provides the molecular gas and dust excitation conditions, offering new insights into the cold ISM properties of this system.

%-------------------------------------- 
   \begin{figure*}
    \includegraphics[width=1.\textwidth]{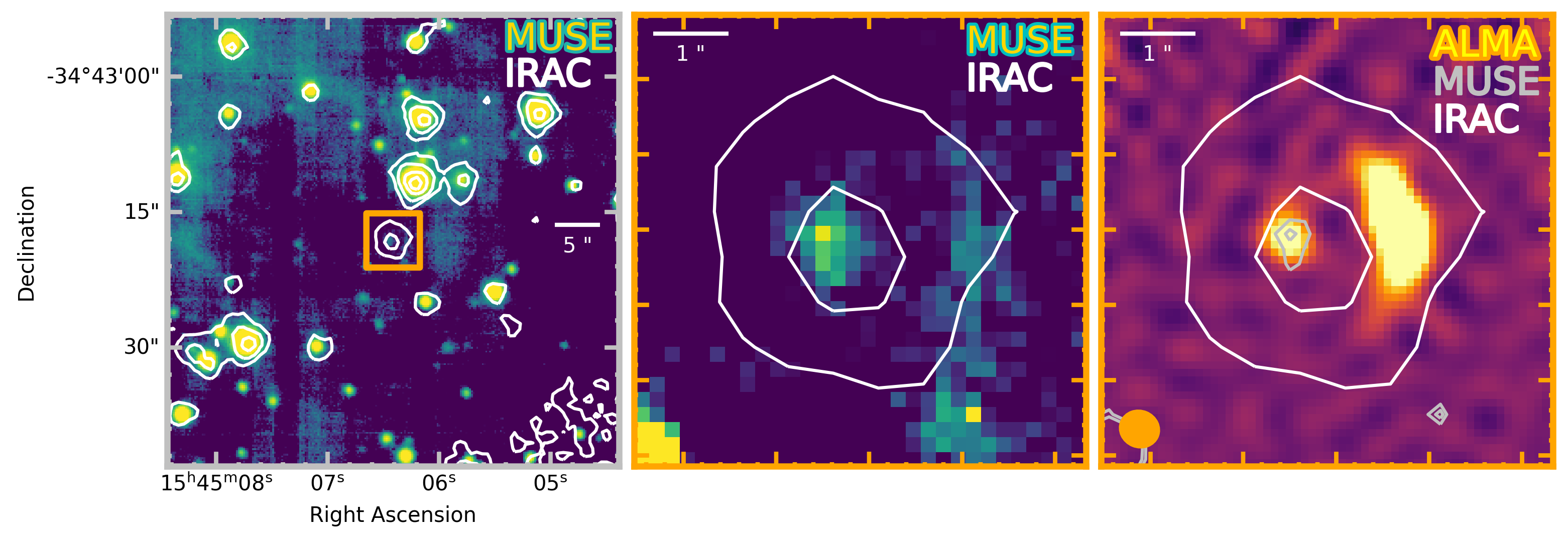}
   \caption{
   \textit{Left panel:} Multi Unit Spectroscopic Explorer (MUSE) white-light (4750 to 9350 $\AA$) median stack image of a  $50\times \ang{;;50}$ region centred on the lens coordinates, with the orange square representing a $3\times \ang{;;3}$ region. The 1, 2, 5, and 10 $\sigma$ Infrared Array Camera (IRAC) contours are shown in white.  
   \textit{Central panel:} $6\times \ang{;;6}$ MUSE zoom-in with the 1 and 2 $\sigma$ IRAC white contours overlaid. 
   Both IRAC and MUSE images are aligned using \textit{Gaia}-DR3 stars within a $\ang{;;30}$ radius circle around the source.
   The faint linear emission feature does not overlap with the ALMA main arc, but instead corresponds to the noise arising from the `gaps' between the MUSE detectors.
   \textit{Right panel:} Same as the central panel, but overlaid on the ALMA Band 7 image of the background galaxy at $\ang{;;0.75}$ resolution (robust=1) . The IRAC and MUSE 1 and 2 $\sigma$ contours are shown in white and grey, respectively, to illustrate the relative position of the lens with respect to the background galaxy in the image plane.
   }
\label{Fig:MUSEdata}
\end{figure*}
%--------------------------------------------------------------------

%%%%%%%%%%%%%%%%%%%%%%%%%%%%%%%%%%%%%%%%%%%%%%%%%%%
\section{Observational data for J154506}
%%%%%%%%%%%%%%%%%%%%%%%%%%%%%%%%%%%%%%%%%%%%%%%%%%%

Star-forming regions such as the Lupus-I molecular cloud are prime targets for multi-wavelength observations, yielding a wealth of data across the electromagnetic spectrum. 
In the following, we describe the data used in this work for the spectroscopic confirmation and characterisation of J154506. 

%--------------------------------------------------------------------
\subsection{ALMA ancillary data}
\label{Sect:ALMA ancillary data}
Extensive archival submm and millimetre data from both the 7 m and 12 m ALMA arrays are available for J154506 (see Table~\ref{table:ALMAdata}).
A more in-depth description of these ALMA data can be found in \cite{SantamariaMiranda2021}; here we summarise the relevant details.
The data reduction, calibration, and imaging of the ALMA Band 3, 6, and 7 (B3, B6, and B7) 
archival data for J154506 were performed using
\textsc{CASA} versions 5.6.1-8, 5.1.1-5, and \textsc{5.6.1-8} \citep[][]{McMullin2007}, respectively. 
We combined all spectral windows to produce both a dust continuum image and a spectral cube using natural weighting for each band. 
We found no significant line detection (i.e. $>3\sigma$) at either the raw or binned spectral resolution (up to $\simeq$100~km\,s$^{-1}$), even after applying uv-tapering to increase the significance. 
This is unsurprising, as these spectral tunings were designed to target brown dwarfs within the Lupus-I molecular cloud and were instead selected to cover local CO transitions. 
As noted above, 
ALMA B6 and B7 continuum observations (IDs: 2015.1.00512.S and 2018.1.00126.S; PI: de Gregorio-Monsalvo, I.) 
provide sub-arcsec resolution (\ang{;;0.75}) images that reveal the existence of 
at least two emitting regions at the coordinates of J154506, located at the edge of the B7 primary beam
(see Figs.~\ref{Fig:MUSEdata} and \ref{Fig:ALMAarchivalData}). 
We re-imaged the B6 and B7 data using \textsc{Briggs} weighting with robust=0.5 and 1. 
These images allowed us to constrain the relative position of the foreground lens and the background galaxy with improved accuracy, 
and were also used to model the system in the image plane (see Section~\ref{Sect:Lens_model_magnification}).

%--------------------------------------------------------------------
\subsection{ALMA Cycle 10 Atacama Compact Array observations}
\label{Sect:ACAobservations}
The Atacama Compact Array (ACA) Cycle 10 spectral scans in Bands 3 and 4 
(2023.1.00251.S; PI: Alcalde Pampliega, B.), 
were conducted to spectroscopically confirm the redshift of J154506. 
The observations consist of six observing blocks (OBs), three in each of Band 3 and Band 4 (B3 and B4), 
and were carried out between November 9, 2023 and  January 11, 2024, 
with a total time on-source range of 9-18 minutes per OB. 
We used a total of 9-11 antennas, 
reaching maximum baselines ranging from 48.0 to 48.9~m and 
an angular resolution of 6-\ang{;;11}.
The six tunings provided continuous coverage 
of two frequency ranges: 90-111 and 139–162~GHz in B3 and B4, respectively.
We observed 12 spectral windows, each of 1.875 GHz, with 15.6 MHz channelisation (28 to 53 km/s). 
Details of the central sky frequencies and sensitivity values are provided in Table~\ref{table:ACAdata}.
Similar spectral setups, which maximise the likelihood of detecting two emission lines, 
have proven very efficient \citep[e.g.][]{Neri2020, Bakx2022} 
in confirming the redshift of dusty high-redshift galaxies through detections of CO, [CI], and even H$_2\rm{O}^+$ emission lines. 

We performed reduction and calibration using the 
CASA-6.5.4-9 \citep[\textsc{}][]{CASA2022} pipeline version, including visual inspections to identify and remove data with irregular amplitude or phase values.
We created continuum images using the interactive \textsc{tclean} task within
\textsc{CASA} with natural weighting. 
Each spectral cube was imaged separately and cleaned 
using a fixed elliptical mask of $\ang{;;16}\times \ang{;;10}$ (major by minor axis) centred at the position of the continuum emission. 
During the data reduction steps, the cubes were further binned to velocity resolutions of $\simeq$100 km/s to maximise the signal-to-noise ratio (S/N) and facilitate line detection. 
To mitigate contamination from potential spectral lines and account for low S/N, 
we initially generated no-continuum-subtracted cubes (see Fig.~\ref{Fig:ACAspectra}). 
We subsequently measured the continuum and extracted it from line-free regions adjacent to detected emission lines.
 Table~\ref{table:ACAdata} lists the characteristics of the final images.

%--------------------------------------------------------------------
\subsection{Large Millimeter Telescope observations}

%-------------------------------------- 
\begin{figure*}
\includegraphics[width=1.\textwidth]{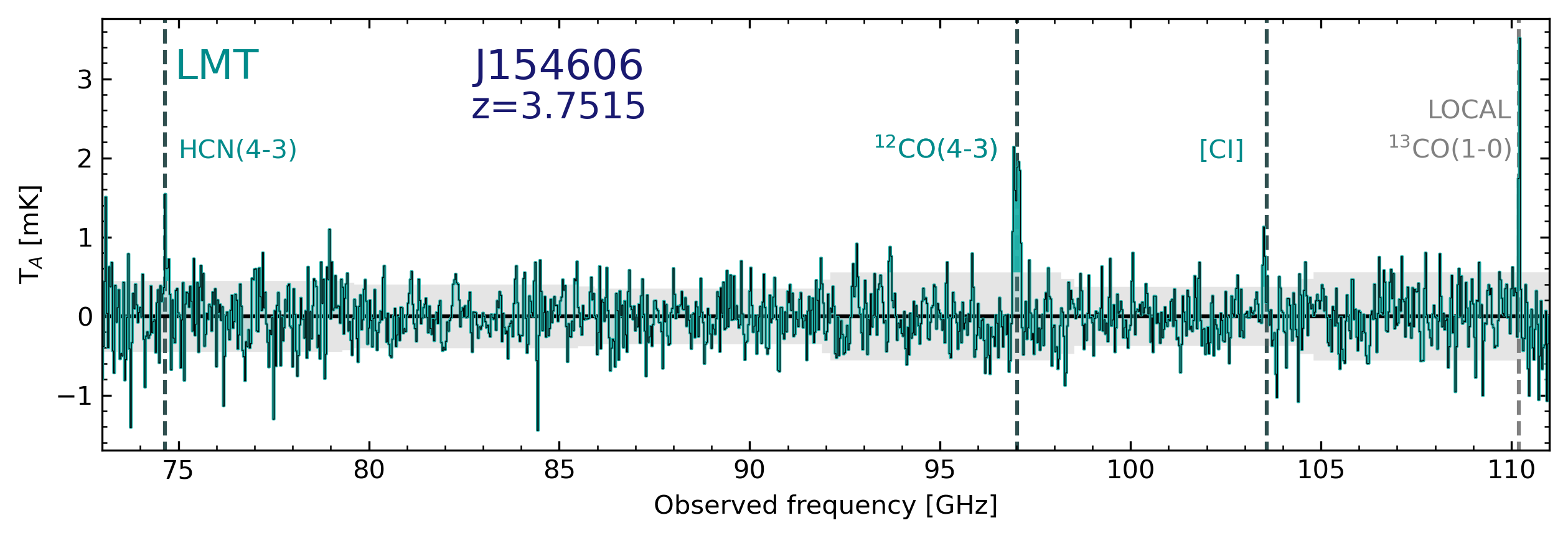}
\includegraphics[width=.5\textwidth]{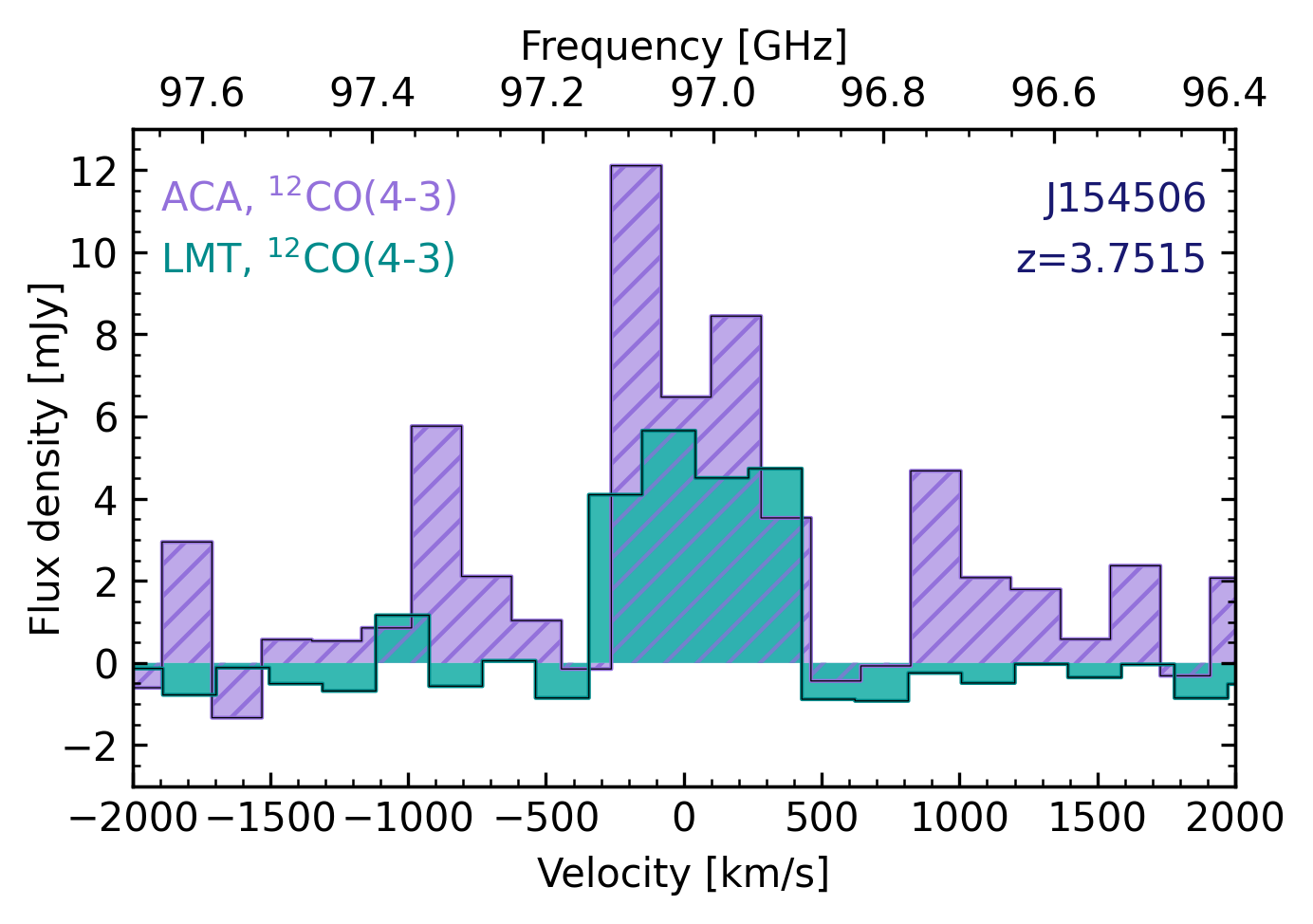}
\includegraphics[width=.5\textwidth]{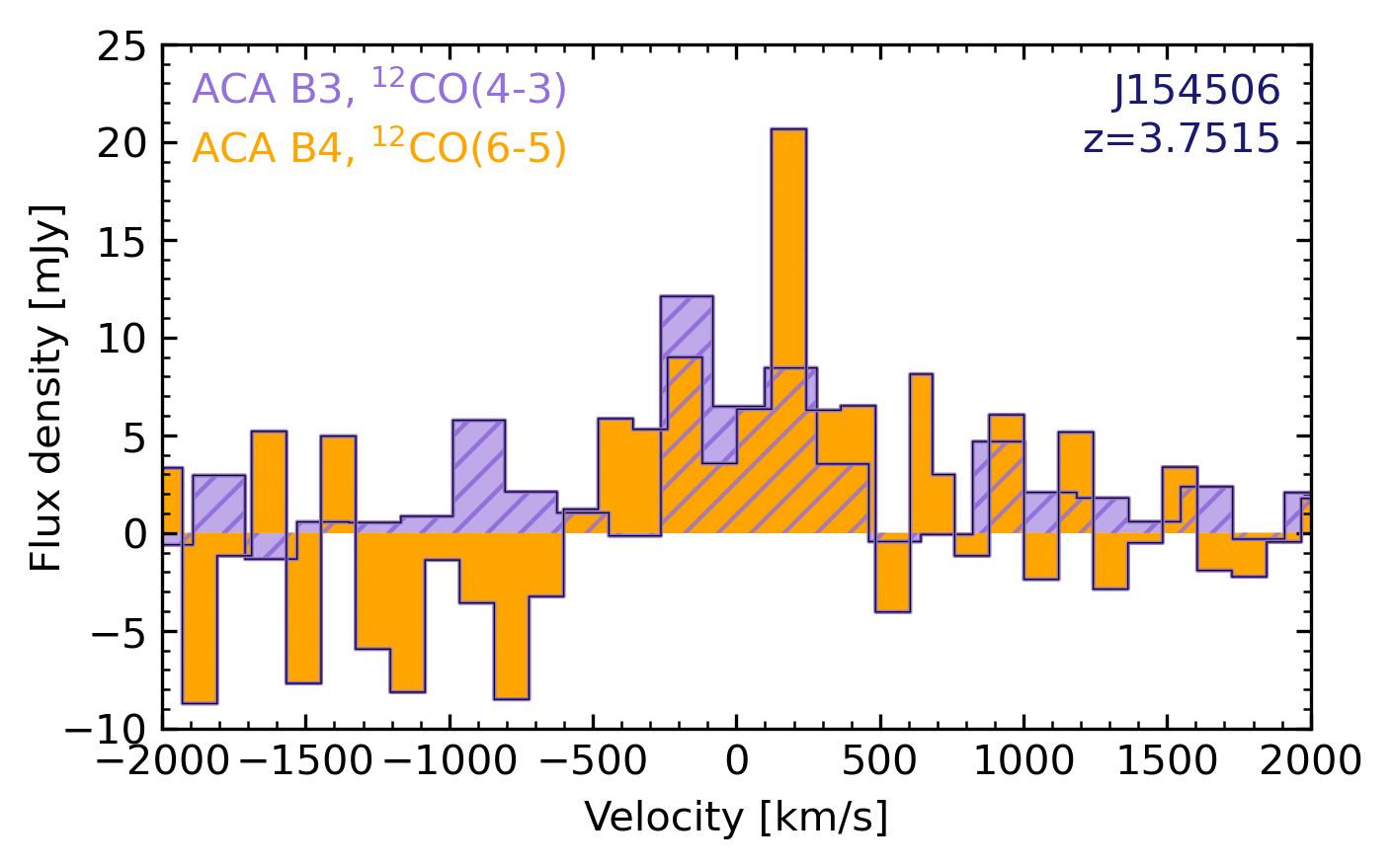}
 \caption{
 \textit{Top panel}: Large Millimetre Telescope
(LMT) spectrum of J154506 obtained with the Redshift Search Receiver (RSR) instrument. 
Transitions with S/N>3 compatible with either the J154506 redshift or the local molecular clouds are marked with vertical dashed lines and labelled in teal and grey, respectively.
\textit{Bottom}: Velocity comparison of the continuum-subtracted line profile for the $^{12}$CO(4-3) line detected with ACA and LMT, respectively (\textit{left}), and $^{12}$CO(4-3) versus $^{12}$CO(6-5) transitions from ALMA/ACA (\textit{right}). The LMT spectrum in the bottom panel is binned to two-channel width to match the velocity resolution of the compared lines.
}
\label{Fig:LMTspectra}%
\end{figure*}
%-------------------------------------- 

We performed 
spectroscopic observations of J154506 with the 
50\,m Large Millimetre Telescope Alfonso Serrano \citep[LMT; ][]{Hughes2010}, 
in the 3 mm band using the Redshift Search Receiver \citep[RSR; ][]{Erickson2007}. 
The LMT/RSR provides simultaneous bandwidth coverage between 73 and 111 GHz
in a single tuning, with an effective beam size of $\simeq$\ang{;;15}.
The full frequency range of the receiver is covered using  six spectrometers, 
with sub-bands overlapping at the band edges 
(73.0, 79.7, 86.0, 92.1, 98.6, 104.9, and 111.0 GHz).
Each band has 256 channels, with a 31.25~MHz channel width, 
corresponding to a velocity resolution of $\simeq$100~$\rm{km~s}^{-1}$ at 3mm.
The observations were carried out on March 4, 2024 under programme 2024-S1-MX-11 (PI: Jimenez-Andrade, E.). 
The total on-source time was 0.9~h, resulting in per-sub-band sensitivities of
1.05, 1.0, 0.94, 1.6, 1.12, and 1.8~mJy rms.
The LMT provides science-ready data products through the standard data reduction process. 
For this dataset, careful data flagging was performed to 
increase the S/N of lower frequency tentative lines. 
To obtain the LMT/RSR results presented here, 
we applied the following gain factor, G:
G$(\nu) = 2.8 \left( \frac{\nu(\text{GHz})}{100 \text{ GHz}} \right) + 0.19 \text{ JyK}^{-1}$.

%--------------------------------------------------------------------
\subsection{APEX observations}

We also performed observations with new FaciLity APEX Submillimeter Heterodyne instrument \citep[nFLASH;][]{Heyminck2006} at the 
Atacama Pathfinder Experiment \citep[APEX; ][]{Gusten2006} 12 m submillimeter telescope. 
The nFLASH receiver has two independent dual-sideband tuneable frequency channels, nFLASH230 and nFLASH460, 
covering intermediate frequency (IF) ranges of 4–12 GHz and 4–8 GHz, respectively, 
allowing simultaneous observations in both channels. 
However, nFLAS460 requires much better atmospheric conditions than nFLASH230
due to the lower atmospheric transmission at higher frequencies. 
Therefore, the observations did not reach the requested rms. 
J154506 was only partially observed under the projects 
C-0113.F-9710C (PI: Alcalde Pampliega, B.) and C-114.F-9703C (PI: Harrington, K.). 
Both projects used nFLASH230 and nFLASH460 tuned to sky frequencies of 217.5 and 400~GHz, 
targeting [CII] and CO(9-8), respectively. 
The total integration times were 10.6 and 63~min (0.5 and 0.2~mK rms) for nFLASH230 and 
10.6, 21.3, and 28.5~min (1.5, 1.7, and 1.5~mK rms) for nFLASH460.
We reduced the APEX data using a consistent strategy, modifying 
the APEX template reduction script using the GREnoble Graphics (GREG) and 
Continuum and Line Analysis Single-dish Software (CLASS) packages within 
Grenoble Image and Line Data Analysis Software (GILDAS)\footnote{https://www.iram.fr/IRAMFR/GILDAS/}.
We smoothed the spectrum from each scan to 40 and 100~km\,s$^{-1}$ channel resolution and
averaged it after subtracting a first-order baseline from the line-free channels of all scans.

%--------------------------------------------------------------------
\subsection{MUSE and FORS2 observations}
Observations with the Multi Unit Spectroscopic Explorer (MUSE) and the FOcal Reducer and low dispersion Spectrograph 2 (FORS2) primarily aimed to confirm the redshift of the foreground lens through the detection of optical emission and absorption lines. Although J154506 was within the MUSE field of view, the probability of detecting its redshifted optical emission lines was very low.
Observations were obtained as part of the ESO filler programmes 111.24UJ.009 (PI: Bian, F.)
with MUSE \citep[]{Bacon2010} 
and 111.24P0.008 (PI: Berton, M.) with FORS2, 
mounted at UT4 and UT1 of ESO’s Very Large Telescope (VLT) at Cerro Paranal in Chile.
The FORS2 spectrograph was used with the GRIS 600z+23 grism and the order separation filter OG590, 
providing a wavelength range of 7400--10000 \AA. 
The FORS2 acquisition was performed through a blind offset, and the observations consisted of 
ten exposures of 2700 seconds each, with the slit width set to $\ang{;;1.3}$.
Due to marginal observing conditions, we could only extract spectra from five datasets. 
Although these were combined, we did not detect spectral features.

The MUSE instrument is an integral field unit spectrograph (R=2000-4000) covering the 4750--9350 $\AA$ wavelength range
with a $\ang{;;0.2}$ spatial resolution across a $60\times \ang{;;60}$ field of view (i.e. the wide field mode). 
The MUSE observations comprised eight observing blocks (OBs), 
each consisting of four exposures of 700 seconds.
We reduced all data using standard procedures with the ESO recipe execution tool (ESOREX).
To further reduce the remaining skylines in the MUSE data, we ran 
the Zurich Atmosphere Purge \citep[ZAP, ][]{Soto2016} (ZAP) sky subtraction tool.
All but two of the MUSE OBs were taken under poor atmospheric conditions (i.e. thick clouds and a seeing $>\ang{;;1}$). 
In the best two OBs, taken on June 21 and 19, 2023, with an airmass $\simeq$1 and a seeing of 0.7 and \ang{;;1.3}, respectively, the lens was detected ($24.5 \pm 0.3$~mag; see. Fig.~\ref{Fig:MUSEdata}) 
but no clear line features were found. 
To extract the spectra, we used a circular aperture with a
$\ang{;;1}$ diameter in the combined image of these two OBs. 
This aperture was selected to maximise the S/N after testing diameters from 0.5 to $\ang{;;2}$. 
The MUSE data also enabled better characterisation of the lensing galaxy position (RA$=$15:45:06.441, DEC$=$-34:43:18.128).
We aligned MUSE cubes using \textit{Gaia}-DR3 stars within a $\ang{;;30}$-radius circle 
around the source, yielding World Coordinate System (WCS) corrections smaller than $\ang{;;0.3}$.
The low quality of this dataset precluded spectroscopic confirmation of the lens and 
prevented further detailed analysis.

%%%%%%%%%%%%%%%%%%%%%%%%%%%%%%%%%%%%%%%%%%%%%%%%%%%
\section{Spectroscopic redshift confirmation}
%%%%%%%%%%%%%%%%%%%%%%%%%%%%%%%%%%%%%%%%%%%%%%%%%%%

%-------------------------------------- 
\begin{figure}
\includegraphics[width=.5\textwidth]{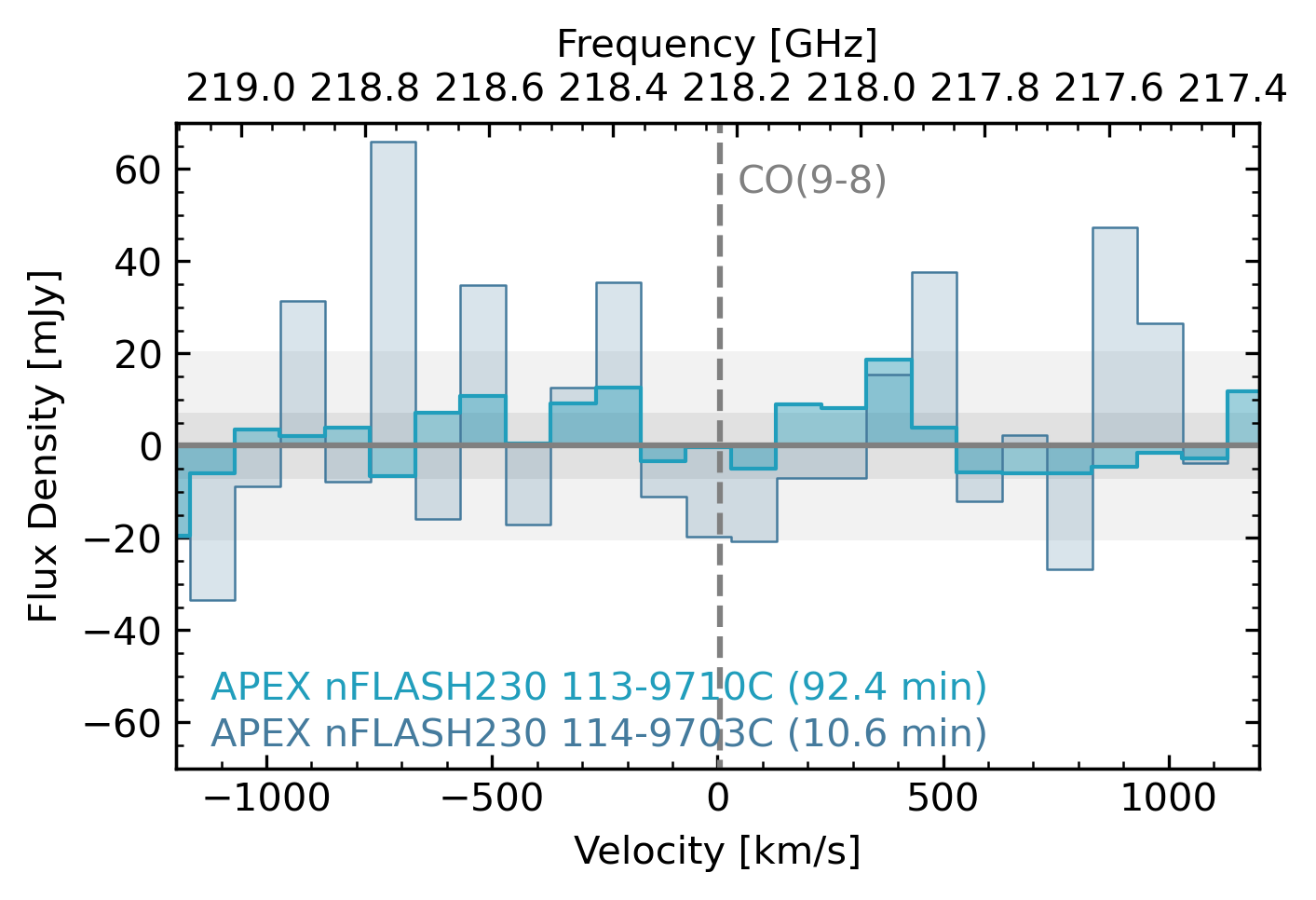}
\includegraphics[width=.5\textwidth]{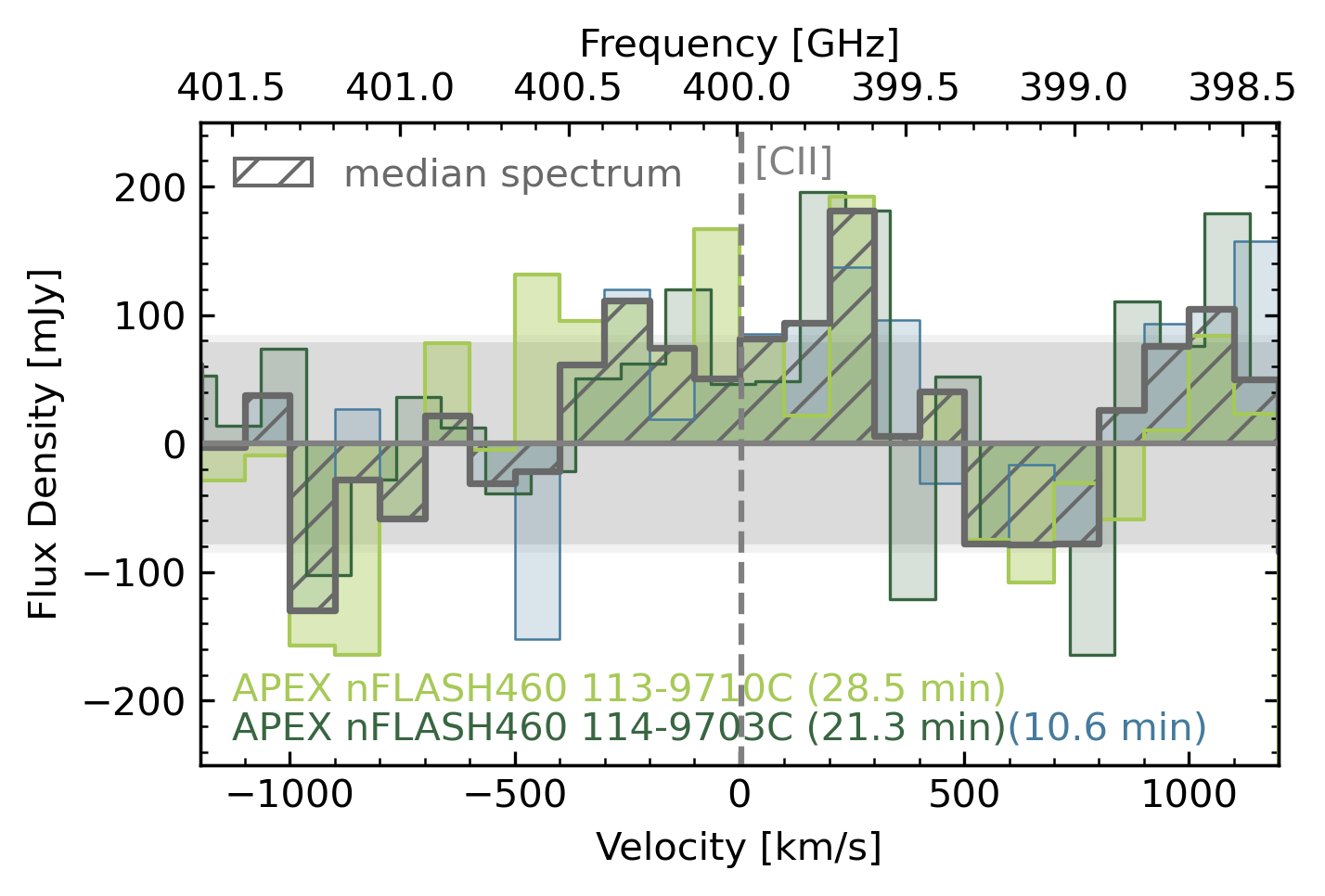}
 \caption{
 Spectra from APEX nFLASH230 (\textit{top panel}) and nFLASH460 (\textit{bottom panel}), centred at the redshifted ($z=3.7515$) frequencies of the CO(9-8) and [CII] emission lines, respectively. 
 The independent runs are colour-coded and labelled accordingly. 
 The derived RMS (1$\sigma$ noise level) is indicated by light grey shading, while dark grey arises from the overlap region of the errors from distinct datasets. 
The CO(9-8) line is not detected, but [CII] is detected at $>2$ sigma in all three scans. 
 For [CII], the median-averaged spectrum (S/N$>$4) is shown in dark grey, with the area shaded using diagonal grey hatching.
}
\label{Fig:APEXspectra}
\end{figure}
%--------------------------------------------------------

%--------------------------------------------------------
\setlength{\tabcolsep}{1pt}
%-------------------------------------- 
\begin{table}[!h]
\caption{
Integrated line fluxes before magnification correction.}             
\label{table:LineFluxes}      
\centering          
\begin{tabular}{ l c c c }     
\hline \hline     
Line  &  Telescope$/$& Line width &Flux density \\ 
           &  Instrument & km s$^{-1}$ &Jy km s$^{-1}$         \\ 
\hline   
$^{12}$CO(9-8)$^{a}$   & APEX/nFLASH & 550 & 2.1 $\pm$  4.3\\
$^{12}$CO(9-8)$^{b}$   & ALMA & 550 & 2.0 $\pm$  1.5\\
$^{12}$CO(6-5)   & ALMA/ACA    &  603 $\pm$ 60 & 6.6 $\pm$ 1.1  \\
$^{12}$CO(4-3)   & ALMA/ACA    &  543 $\pm$ 90 & 5.7 $\pm$ 1.0  \\ 
$^{12}$CO(4-3)   & LMT/RSR     &  580 $\pm$ 97 & 4.0 $\pm$ 0.5  \\
$^{12}$CO(4-3)$^{c}$   & NRO     &  625 $\pm$ 120 & 6.5 $\pm$ 0.7  \\
$^{12}$CO(2-1)$^{c}$   & ATCA     &  543 $\pm$ 106 & 3.0 $\pm$ 0.5  \\
$\rm[CI]$        & LMT/RSR     & 362 $\pm$ 135 & 1.0 $\pm$ 0.3  \\
$\rm[CI]$        & ALMA/ACA    & 340 $\pm$ 113 & 2.9 $\pm$ 1.0  \\
$\rm[CII]^{d}$   & APEX/nFLASH & 598 $\pm$ 85 & 71.5 $\pm$  15.4\\
H$_2\rm{O}^+(2_{02}-1_{11})_{(3/2-1/2)}$   & ALMA/ACA    & 230 $\pm$ 116 & 4.3 $\pm$ 1.0  \\ 
H$_2\rm{O}^+(2_{11}-2_{02})_{(3/2-3/2)}$   & ALMA/ACA    & 548 $\pm$ 110 & 7.5 $\pm$ 1.6  \\  
HCN(4-3)         & LMT/RSR     & 502 $\pm$ 126 & 1.0 $\pm$ 0.3  \\
\hline  \hline                         
\end{tabular}
$^{a}$Value obtained by forcing a measurement in the spectra, assuming a line velocity of 550 km s$^{-1}$.\\
$^{b}$Flux derived from ALMA spectra in which only half of the line is covered. A conservative 75\% error is assigned to the value.\\
$^{c}$Measurements from \cite{Tamura2025}.\\
$^{d}$The provided flux corresponds to the median spectrum; the measurements for the three independent spectra are consistent within the errors.\\
\end{table}
%--------------------------------------------------------

ALMA B6 and B7 sub-arcsec resolution (\ang{;;0.75}) continuum archival observations 
(detailed in section~\ref{Sect:ALMA ancillary data}) detected J154506 at the edge of the ALMA primary beam. 
These images resolved the continuum 
emission into two emitting regions: a point-like detection and a bright arc-like elongated shape (see the right panel of Fig~\ref{Fig:MUSEdata} and Fig.~\ref{Fig:ALMAarchivalData}), 
confirming the gravitational lensing of this extremely bright SMG.

Our ALMA Cycle 10 B3 and B4 spectral scans successfully detected two S/N$\gtrsim$5 emission lines, at 97 and 145.5~GHz (see Fig.~\ref{Fig:ACAspectra} and bottom panels in Fig.~\ref{Fig:LMTspectra}), which is typically sufficient to confirm the redshift of a galaxy. 
However, the degeneracy in CO transitions can lead to ambiguous solutions in certain scenarios. 
In simple terms, multiplying the frequencies of both CO line transitions by the same integer factor can produce another valid transition, resulting in multiple possible redshifts \citep[e.g.][]{Bakx2022}. In this case, the lines could correspond to redshifts of 1.38, 3.75, 6.1, or 8.5. 
Although the ACA scans covered higher-J transitions for the potential redshifts z=6.12 and 8.5, the low S/N of the spectra and the expected faintness of those transitions meant that their non-detection could not confirm the redshift.
The 97 and 154 GHz lines are also compatible with the detection of CO(2-1) and CO(3-2) at $z=1.38$; 
however, we discard this low redshift solution due to its incompatibility with the photometric redshift ($z_{\rm{phot}}>3$). 
Moreover, the derived dust temperature from FIR fitting at $z=1.38$, T$=$15~K is low and unlikely for galaxies at that redshift \citep[e.g.][]{Schreiber2018}. In addition, the intrinsic SFR$_{\rm{IR}}$ would be below $100 \rm{M}_{\odot} \rm{yr}^{-1}$. 

The LMT/RSR spectrum simultaneously covers the 73-111~GHz frequency range, 
overlapping with the ALMA B3 scans (see Fig.~\ref{Fig:LMTspectra}). 
In this frequency range, we expected the lower J CO(5-4) and CO(7-6) transitions for the $z=6.12$ and $8.5$ solutions, respectively. 
In this case, the non-detection 
disfavours
the higher-z solutions. 
We detect the 97~GHz line, further confirming the ALMA detection 
and the spectroscopic redshift of $z_{\rm{spec}}$=3.7515. 

With $z_{\rm{spec}}=3.75$ confirmed, the 97.03 and 145.5~GHz lines correspond to CO(4-3) and CO(6-5), respectively. 
For CO (4-3), the measured line flux is consistent between the two instruments. 
To characterise the velocity-integrated flux density, and
given the non-Gaussian shape of the line profiles, 
we calculated the full line-width at zero intensity (FWZI).
Specifically, we measured an integrated line flux in the sky-plane (lensing-uncorrected) of $5.7 \pm 1.0$ and $4.0 \pm 0.3~\rm{Jy~km~s}^{-1}$ for ALMA and LMT, respectively, over $\sim 550\rm{km s}^{-1}$.
For the higher J transition CO(6-5), 
we obtain an integrated flux of $6.6 \pm 1.1~\rm{Jy~km~s}^{-1}$.
We also find two low S/N (i.e. 4>S/N>2; see Fig.~\ref{Fig:LMTspectra_pannels}) tentative detections at 74.6 and 103.6~GHz, 
consistent with HCN~(J=4-3) and [CI]$^{3}$P$_{1}$-$^{3}$P$_{0}$ at that redshift.

We also find two S/N$\sim3$ detections corresponding to H$_2\rm{O}^+(2_{02}-1_{11})_{(3/2-1/2)}$ and 
H$_2\rm{O}^+(2_{11}-2_{02})_{(3/2-3/2)}$ at 151.9 and 161.3 GHz, respectively.   
Although these specific transitions have not previously been detected in local ultraluminous IR galaxies \citep[e.g.][]{vanderWerf2010,Rangwala2011} or in high redshift SMGs \citep{Reuter2023}, 
the detection of similar rotational transitions at $z=2-4$ \citep{Yang2016, Yang2019} provides compelling support for our detections.
Finally, a very narrow spike is observed at 110~GHz, which most likely corresponds to 
foreground $^{13}$CO(1-0) emission (i.e. from the Milky Way in the direction of J154506).

Observations with APEX covered the CO(9-8) and [CII] lines at $z=3.75$.
Although the project was not completed and CO(9-8) is not detected, 
we find 
an S/N$>$4 detection 
at 400~GHz corresponding to [CII] at $z=3.75$.
Figure~\ref{Fig:APEXspectra} shows the apparent flux density centred at the redshifted 
frequency of the CO(9-8) (218.23~GHz) and [CII] (399.99~GHz) emission lines 
at $\approx100\rm{km\,s}^{-1}$ resolution. 
We did not detect any CO(9-8) emission in the APEX spectra, even at 7~mJy rms. 
However, to constrain the CO Spectral Line Energy Distribution (SLED), 
see Section~\ref{Sect:GasProperties}), we performed a forced measurement 
by integrating the spectrum over a velocity width of 550 km s$^{-1}$ (see Table~\ref{table:LineFluxes}).
Similarly, for the ALMA CO(9-8) spectrum presented in \cite{Tamura2025}, where only approximately half of the line profile is covered and no measurements are provided, we estimated the total flux by scaling the integrated observed spectrum to an assumed line width of 550 km s$^{-1}$.
Due to the inherent uncertainty of such an estimate, we assign a conservative 75\% error to this value. 
The APEX non-detection is consistent with this estimate, as its significantly higher rms is more than three times the ALMA measured flux density.
The three independent[CII] observations are shown in the lower panel of Fig.~\ref{Fig:APEXspectra}. 
Despite their diverse integration times, a very similar sensitivity (i.e. $\approx$80~mJy) was achieved in all cases. 
Although each independent spectrum provides only a S/N$=$2-3 [CII] detection, the consistency between the three runs, along with the very similar widths (see Fig.~\ref{Fig:APEXspectra}) and spectral shape resembling that of CO(6-5), allow us to confirm the presence of [CII] emission. Additionally, the median spectrum yields an S/N $>$4 detection.

%--------------------------------------------------------------------
\begin{figure*}
    \centering
    \includegraphics[width=.9\textwidth]{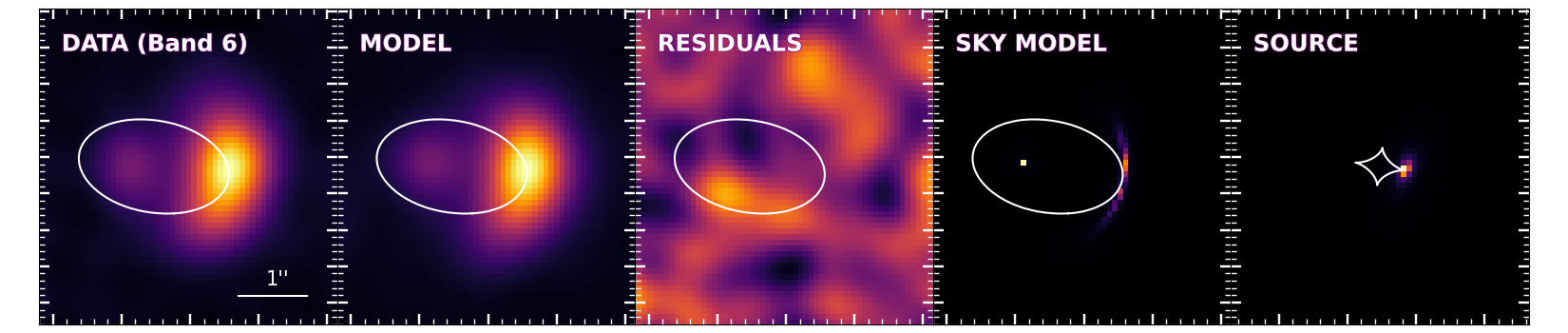}
    \includegraphics[width=.9\textwidth]{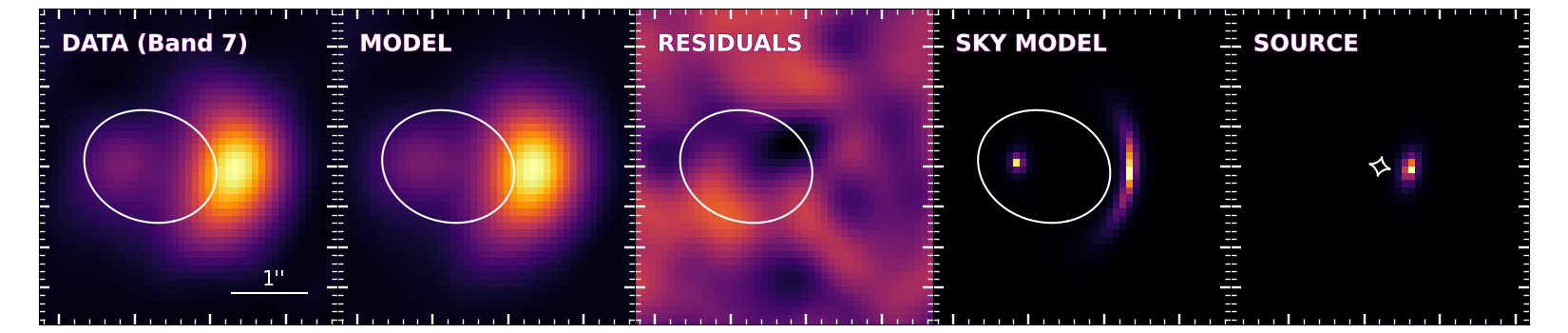}
    \caption{
    Results from the lens modelling analysis in the uv-plane for B6 (\textit{top}) and B7 (\textit{bottom}). From left to right: the panels show the `dirty' data image, `dirty' model image, `dirty' residuals, sky model (i.e. deconvolved), and source model. The white line represents the critical curve (first four panels) and the caustic curve (rightmost panel).
    }
    \label{Fig:Model}
\end{figure*}
%--------------------------------------------------------------------

Our spectroscopic redshift of z$_{\rm{spec}}=3.7515 \pm 0.0005$ is in excellent agreement with the redshift recently reported by \cite{Tamura2025} ($z = 3.753 \pm 0.001$).  Furthermore, our independent ALMA detection of the CO(4-3) transition at 97 GHz yields an integrated line flux of $5.7 \pm 1.0 ~\rm{Jy~km~s}^{-1}$, 
consistent within uncertainties, with their measurement ( $6.5 \pm 0.7~\rm{Jy~km~s}^{-1}$). This agreement 
strengthens the robustness of our flux measurements and derived properties 
(see Sections~\ref{Sect:DustProperties} and ~\ref{Sect:GasProperties}).
The lower RSR/LMT fluxes ($4.0 \pm 0.3 ~\rm{Jy~km~s}^{-1}$ for CO(4-3)) could be influenced by known instrumental systematic calibration biases. \footnote{\href{http://lmtgtm.org/wp-content/uploads/2022/09/RSR_v2p0_reduced.pdf}{LMT-SCI-MAN-002, Version v2.0}}

The observed extreme S$_{250\mu\rm{m}}$/S$_{500\mu\rm{m}}$ flux ratio (see Figure~\ref{Fig:FIRcolours}) and the photometric redshift of J154506 initially suggested a potentially higher-redshift system. 
However, our spectroscopic redshift of $z_{\rm{spec}}$=3.7515, combined with a relatively cold dust temperature (T$=38$~K; see also Section~\ref{Sect:GasAndDustProperties}) reveals a less extreme, though still high-redshift system. 
This highlights the inherent degeneracy between dust temperature and redshift in broadband colour photometric redshift indicators \citep[e.g.][]{Chapman2004, Casey2014, Dowell2014}, emphasising the need for spectroscopic confirmation and comprehensive FIR fitting to characterise dusty star-forming galaxies (DSFGs).
Analysis of the UV-to-Radio SED  often provides more accurate photometric redshifts and physical properties. However, for J154506, foreground contamination combined with the large uncertainty in the Galactic extinction caused by the Lupus-I molecular cloud \citep[see][]{Rygl2013}, makes it challenging to constrain the background source stellar emission. 
The foreground is difficult to model without 
high-resolution, long-integration optical-to-MIR data, whereas 
the Lupus-I extinction 
differentially impacts the SED, causing significant absorption at shorter (UV-to-NIR) wavelengths, 
where it combines with intrinsic galaxy dust extinction. 
These factors render a unified rest-frame UV-to-FIR energy-balanced SED fitting 
unfeasible and makes SED-derived properties, such as the stellar mass, highly uncertain.

%%%%%%%%%%%%%%%%%%%%%%%%%%%%%%%%%%%%%%%%%%%%%%%%%%%
\section{Lens modelling and magnification}
%%%%%%%%%%%%%%%%%%%%%%%%%%%%%%%%%%%%%%%%%%%%%%%%%%%
\label{Sect:Lens_model_magnification}

Deriving the intrinsic physical properties of J154506 requires both a magnification factor and a confirmed redshift. To determine the magnification factor, we constructed a mass model for the lens and a light model for the source that best fit our observations. 
Because the spectroscopic redshift of the foreground lens could not be constrained, 
our modelling primarily relied on reproducing the observed image geometry, 
which is less sensitive to the exact lens redshift.
The magnification factor, which is primarily dependent on the image geometry, remains largely unaffected by the unconstrained lens redshift.
For this analysis, we used the publicly available software {\sc PyAutoLens} \citep{ Nightingale2015, Nightingale2018} and conducted the analysis directly in the uv-plane \citep[e.g.][]{Dye2018, Maresca2022}. 
This uv-plane modelling approach consistently outperforms sky-plane modelling for simulated ALMA observations by avoiding biases and spurious structures introduced by the \textsc{CLEAN} process \citep{Maresca2025}.
We modelled the lens as a spherical isothermal ellipsoid (SIE) and the source as a Sérsic profile \citep[][]{Sersic1963}. 
The model parameters were optimised by fitting only the continuum emission in B6 and B7 independently, using the non-linear sampler Dynesty. Note that for B6, we used only spectral windows 0 and 1 due to the presence of emission lines in the other spectral windows.
The lens centre was left as a free parameter, with a Gaussian prior centred between the two multiple images and a standard deviation of \ang{;;1}. 
The lens position recovered by the parametric model is consistent with the centroid of the MUSE detection. The median positional offset (\ang{;;0.1}) is better than the MUSE spatial sampling (\ang{;;0.2}/pixel).
The best-fit lens mass model parameters are $R_{\rm E} = 0.86 \pm 0.01$, $q = 0.58 \pm 0.02$, and $\theta = -8 \pm 2$, in B6, and $R_{\rm E} = 0.78 \pm 0.01$, $q = 0.81 \pm 0.03$, and $\theta = -10 \pm 7$, in B7.
Here, $R_{\rm E}$ is the Einstein radius (arcsec), $q$ is the axis ratio defined as the ratio of the minor axis to the major axis ($q = b/a$), and $\theta$ is the position angle (in degrees) of the major axis measured counterclockwise from the positive x axis in the image plane coordinates.
The derived background-source parameters for B7 are an effective radius of 
$R_{\text{eff, B7}} = 0.147^{+0.009}_{-0.008}$ 
arcsec and a source centroid position relative 
to the lens centre of $(x, y)_{\text{B7}} = (0.56\pm +0.01, -0.07\pm 0.01$ arcsec. For B6, we find an effective radius of $R_{\text{eff, B6}} 
= 0.22^{+0.12}_{-0.04}$ arcsec and a source centroid position of $(x, y)_{\text{B6}} = (0.49\pm 0.01, -0.01\pm+0.02)$ arcsec. The larger and more asymmetric uncertainty in $R_{\text{eff, B6}}$ reflects the slightly lower S/N relative to B7.
Figure~\ref{Fig:Model} presents the results of the modelling analysis. Given the low resolution of our data, the simple parametric models used for the lens and source are sufficient to fit the observations down to the noise level. 
However, a subtle correlation between the residual images in B6 and B7 suggests that the model may not fully capture the intricate structure of the source or could hint at unmodelled substructure within the foreground lensing object. 
We derive magnification factors of $\mu = 7.4 \pm 0.6$ and $\mu = 4.5 \pm 0.4$ for B6 and B7, respectively. For the remainder of this paper, we adopt the average magnification $\mu = 6.0 \pm 0.4$ to derive intrinsic properties.
Our derived magnification factors agree with those recently reported by \citep[$\mu = 6.6$;][]{Tamura2025} from a simple model in the image plane assuming $z_{\rm{lens}}=0.5$

%%%%%%%%%%%%%%%%%%%%%%%%%%%%%%%%%%%%%%%%%%%%%%%%%%%
\section{Properties of the cold interstellar medium}
%%%%%%%%%%%%%%%%%%%%%%%%%%%%%%%%%%%%%%%%%%%%%%%%%%%
\label{Sect:GasAndDustProperties}

In this section, we derive the initial estimates of the cold interstellar medium properties 
by modelling the measured dust photometry and velocity-integrated line flux densities. 
First, we analyse the well-sampled FIR SED of J154506 to characterise 
its dust continuum emission and derive global FIR properties.
Next, we present the results of our combined dust and line modelling 
using the TUrbulent Non-Equilibrium Radiative transfer (\tuner) framework (See Section~\ref{Sect:GasProperties} and Appendix~\ref{App:CO_SLED}). 
Finally, we discuss [CII]-derived properties and 
the potential contribution of an AGN.

% %%%%%%%%%%%%%%%%%%%
\subsection{Dust emission and FIR-derived properties}
\label{Sect:DustProperties}
%%%%%%%%%%%%%%%%%%%%%%%%%%%%%%%%%%%%%%%%%%%%%%%%%%%
%-------------------------------------- 
\begin{figure}
    \includegraphics[width=0.5\textwidth]{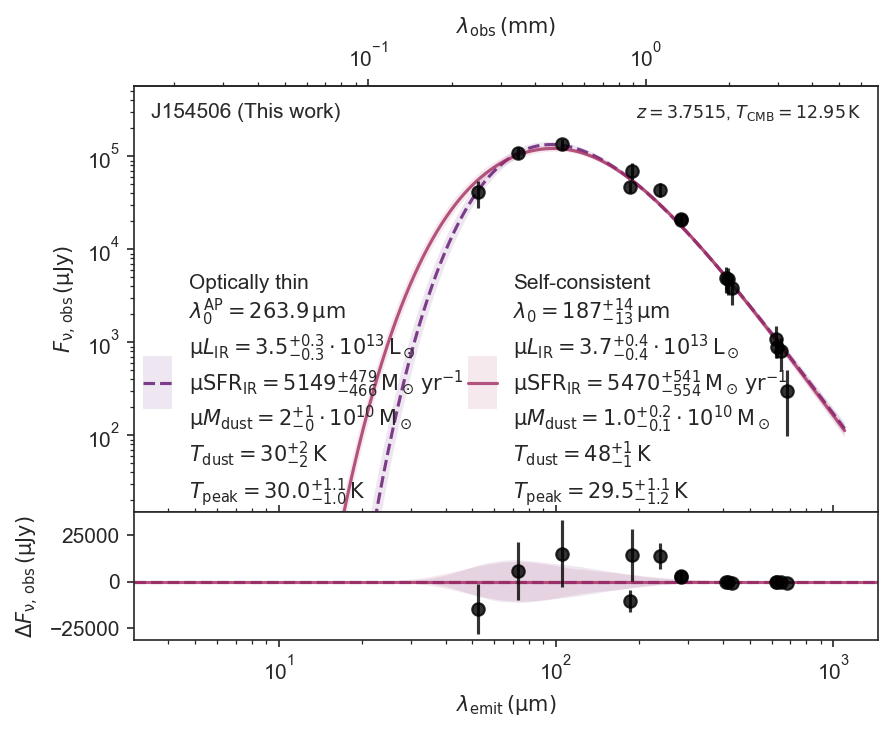}
   \caption{Far-infrared SED of J154506 fitted using \textsc{MERCURIUS}. The black circles indicate the dust continuum detections.
   The dashed line shows the best-fit model assuming optically thin emission. The main derived physical properties, before magnification correction, are also shown. The corner plot and magnification-corrected values are provided in Appendix~\ref{App:Dust}. 
 }
    \label{Fig:MERCURIUSfit}
\end{figure}
%-------

We first fitted J154506 using the Bayesian FIR SED PYMULTINEST
fitting code \textsc{MERCURIUS} \citep[Multimodal Estimation Routine for 
the Cosmological Unravelling of Rest-frame Infrared Uniformised Spectra; ][]{Witstok2022}, 
which employs a nested sampling algorithm \citep[\textsc{PYMULTINEST, }][]{Buchner2014} 
to fit a greybody spectrum to FIR photometry.  \textsc{MERCURIUS} explicitly accounts for the cosmic microwave background (CMB)  following the approach of \cite{daCunha2013}.
For fitting, we considered rest-frame FIR wavelengths from 10 to 10$^{3}\mu$m.
First, we assumed an entirely optically thin scenario for dust emission, and 
left the dust temperature (T$_{\rm{dust}}$) and emissivity index ($\beta_{\rm dust}$) free to vary.
We then used a more physically realistic, self-consistent dust opacity model 
that accounts for wavelength dependence and the transition between optically thin and thick regimes, 
parametrised following \cite{Witstok2022}.
We set our T$_{\rm{dust}}$ priors by disfavouring extremely high dust temperatures, consistent with recent studies \citep[e.g. ][]{Witstok2022, Witstok2023, Valentino2024}. 
Specifically, we used a default gamma distribution with a shape parameter (a=1.5), and 
for $\beta_{\rm dust}$, we imposed a Gaussian prior centred at 1.8 with a  standard deviation of 0.25.
Following \cite{Schouws2022b}, we adopted a dust emissivity coefficient ($\kappa$) of the form $\kappa(\nu) = \kappa(\nu$/ $\nu_{0})^{\beta}$, with $\kappa = 8.94~\rm{cm}^{2}\rm{g}^{-1}$
at $\lambda_{\rm{ref}} =  158 \mu$m \citep{Hirashita2014}.
This value of $\kappa$ is appropriate for dust grains ejected by supernovae (SNe) after reverse shock destruction \citep{Hirashita2014}.

Fig.~\ref{Fig:MERCURIUSfit} presents the best fit of a modified blackbody for J154506. 
Briefly, we obtain an emissivity indices of $\beta_{\rm dust}$ $=2.0 \pm0.1$ and  $2.1 \pm0.1$, 
and dust temperatures of $30\pm2$ and $48\pm1$~K for the optically-thin and self-consistent scenarios, respectively.
The derived intrinsic IR luminosity is L$_{\rm{IR}}=6-7\times 10^{12}~\rm{L}_{\odot}$,
and the IR-based SFR ranges from $860\pm80$ to $910\pm90~\rm{M}_{\odot}yr^{-1}$.  
Additionally, after correcting for magnification, we derive dust masses of $3.4\pm1.6\times10^{9}~\rm{M}_{\odot}$ and $1.7\pm0.3\times10^{9}~\rm{M}_{\odot}$ for the optically-thin and variable dust optical depth models, respectively.

The most robust quantities in our analysis are the integrated IR luminosities (L$_{\rm{IR}}$ and L$_{\rm{FIR}}$) and the approximate peak of the dust SED, which are very well constrained by all available integrated flux measurements across the dust SED.
While this approach provides a useful first-order characterisation, several parameter degeneracies are inherent to such simplified, integrated modelling.
Dust mass estimates are highly dependent on the assumed $\kappa$ value and optical depth, which can vary with grain composition and processing history. For example, if J154506 has a dominant contribution from dust grown in the ISM (e.g. graphite and silicate grains) with higher $\kappa$, this could lead to an overestimation of the inferred dust mass. 
The fitted T$_{\rm{dust}}$ can also degenerate with $\lambda_0$ ($\lambda$ at which the optical depth equals unity), $\beta$, and the assumed optical depth regime. 
In particular, adopting an optically thin assumption—where radiation from a central source passes through dust without significant absorption or re-emission—is likely unrealistic for a highly dust-obscured system such as J154506, as many high-redshift starbursts show substantial FIR optical depths \citep[e.g.][]{Jin2022}. Nevertheless, even the self-consistent model employed here contains simplifying assumptions, including neglecting multiple dust components, temperature distributions along the line of sight, and interactions with the surrounding gas.
The variation in dust temperature predictions across models underscores the 
limitations of oversimplified radiative transfer approaches and the need for a more sophisticated modelling (see Section~\ref{Sect:GasProperties}).

% %%%%%%%%%%%%%%%%%%%
\subsection{Radiative transfer\ model results}
% %%%%%%%%%%%%%%%%%%%
\label{Sect:GasProperties}

We employed the state-of-the-art \tuner model 
to simultaneously fit the dust and CO line SEDs and to derive the molecular gas excitation properties. 
The \tuner\, model is described in detail by \cite{Harrington2021}, and we refer the reader to that work 
for a comprehensive overview \citep[see also ][]{Strandet2017,Jarugula2021}. 
Briefly, these non-local thermodynamic equilibrium (non-LTE) radiative transfer calculations model the CO line intensities and dust continuum
using a large velocity gradient approximation \citep[LVG; see, e.g. ][]{GoldreichKwan1974} 
and a lognormal probability distribution function 
to describe the gas volume density \citep[see ][]{Krumholz2005}. 
For this initial combined line and continuum SED analysis, 
we followed \cite{Harrington2021} and fixed some of the unknown parameters. 
Specifically, we fixed the [CO]/[H$_2$] gas-phase abundance 
to a fiducial Milky Way value (i.e. log([CO]/[H$_2$]=-4.0).
The initial model fits, obtained with a free molecular gas-to-dust mass ratio (GDMR), 
yielded a median GDMR of $100$, which we subsequently fixed to reduce degeneracy in the parameter space.
Additional details are described in Appendix \ref{App:CO_SLED}. 
We note that in this modelling we explicitly assumed the same value 
of $\kappa = 8.94~\rm{cm}^{2}\rm{g}^{-1}$ at $158 \mu$m as used in the MERCURIUS modelling above. \\

%--------------------------------------------------------------------
\begin{figure*}[!h]
    \centering
    \includegraphics[width=.9\textwidth]{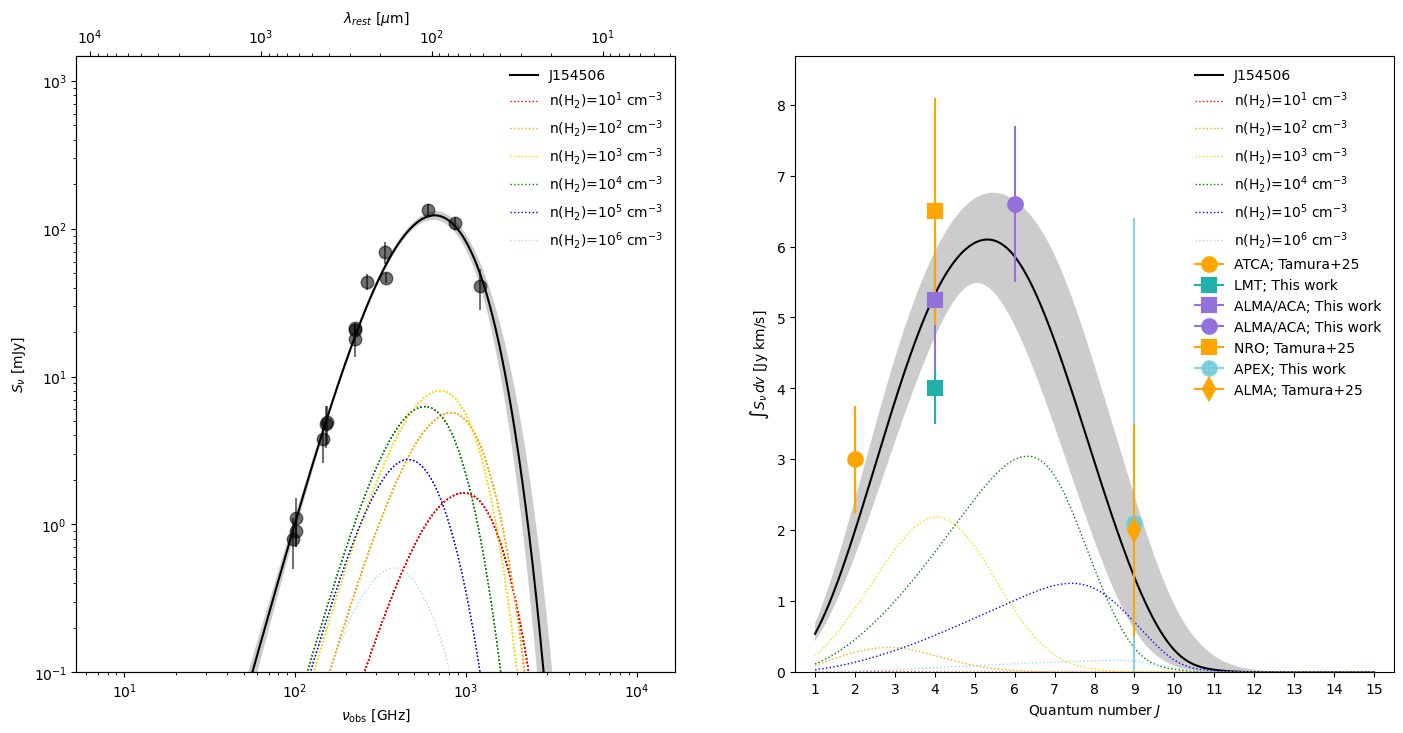}
    \caption{
    \label{fig:dustcosed}
   Results from the \textit{\tuner} model for the observed dust SED and velocity-integrated CO line flux densities. The 50th percentile is shown by a dashed black line and the 16-84th percentile range is shown as a grey shaded area. The different dashed, colour-coded curves denote the representative contributions to the observed line and continuum SEDs from gas at specific molecular gas densities. 
    We sampled the density grid using 50 bins, but we only plot a subset of representative density components (coloured lines) to highlight their relative contribution.
    These densities correspond to log(n(H$_2$)) = 1 (red), 2 (orange), 3 (yellow), 4 (green), 5 (dark blue), and 6 (light blue) cm$^{-3}$, as determined by sampling the lognormal probability distribution function of the mean \Htwo\  density. 
    For the CO SED (right panel), contributions are scaled up by a factor of five for visibility.
    All observed data are shown as colour-coded symbols.
    }
\end{figure*}
%--------------------------------------------------------------------

Figure \ref{fig:dustcosed} presents the best-fit model along with the 16th-84th percentile ranges 
for the observed CO line and dust SEDs. 
The lognormal distribution derived from the best-fit mean density 
is sampled by 50 bins spanning densities of 10-10$^{7} \rm{cm}^{-3}$, 
as shown in Figure \ref{fig:dustcosed}. 
The dust SED peak and turnover are well-sampled, 
with dominant contributions from more diffuse molecular gas 
in the 10$^{2-4}$ cm$^{-3}$ range. 
This suggests that most of the FIR emission originates from 
this relatively diffuse to moderately dense molecular medium, 
rather than from highly compact star-forming regions 
that may be more common in extreme starbursts or AGN-dominated systems 
\citep[][]{Narayanan2014, Scoville2017, Harrington2021}.
The CO line SED peaks near the CO(5-4) transition, 
exhibiting a sharp turnover that contrasts with the flatter high-J CO SLEDs 
of quasi-stellar objects (QSOs) and the most extreme starbursts
\citep[e.g.][]{Rosenberg2015, Harrington2021}. 
Overall, CO line excitation is dominated by slightly denser molecular gas of 10$^{3-4}\rm{cm}^{-3}$, 
which is typical of main-sequence 
DSFGs and 
moderate starbursts at high redshift \citep[][]{Daddi2015, Valentino2020, Liu2021}. 

The derived dust properties are consistent with the Ultra-Luminous InfraRed Galaxy (ULIRG) nature of this object, 
with a 16th-84th percentile range of 
L$_{\rm{IR}}(8-1000 \mu \rm{m})=5.5-6.5\times10^{12} \rm{L}_{\odot}$ after correcting for magnification, 
consistent with the MERCURIUS fits.
The well-constrained \tuner\ model fits the observed data comparably to the best-fit \textsc{MERCURIUS} model, yielding a dust emissivity index of $2.2$ and a relatively cold dust temperature of $38\,\rm{K}$. 
This finding supports the prevalence of cold dust in high-redshift starbursts 
as a natural consequence of deeply embedded SF within gas- and dust-rich environments \cite[e.g.][]{Scoville2023}, 
a phenomenon first predicted by early dust SED models for young stars \cite[e.g.][]{Scoville1976}. 

Understanding the coupling between gas and dust in galaxies provides crucial insights into 
the dominant heating mechanisms and overall physical conditions of their ISM.
A key diagnostic for this coupling is the ratio of the gas kinetic temperature (T$_{\rm{k}}$)
to the dust temperature (T$_{\rm{d}}$), since distinct heating processes affect these components differently.
Although J154506 has a strong intrinsic far-IR luminosity, 
the median gas-to-dust temperature ratio is T$_{\rm{k}}$/T$_{\rm{d}} =1.1 \pm 0.2$. 
The median derived value for similarly bright $z=1-3.5$ \textit{Planck}-selected starbursts 
is T$_{\rm{k}}$/T$_{\rm{d}}$ = 2-3 \citep{Harrington2021}. 
This ratio has been proposed as a proxy for mechanical versus photoelectric heating contributions
\citep{Harrington2021, Dunne2022} that may be responsible for molecular gas heating. 
The near-unity value for J154506 is considerably lower, suggesting that mechanical heating is not a dominant powering mechanism. 
As discussed by \citet{Harrington2021}, this value approaches unity 
when the gas and dust temperatures are coupled, 
typically at mean densities above 10$^{4-5}\rm{cm}^{-3}$. 
The 10$^{4}\rm{cm}^{-3}$ CO-emitting gas dominates both the observed CO line SED and the molecular gas mass in this system. 

The model fit parameters enable the calculation of the total intrinsic molecular gas mass, M$_{\rm{ISM}} = 3.1-8.5 \times 10^{11} ~\rm{M}_{\odot}$, CO luminosity L$^{\prime}_{\rm{CO(1-0)}}=4.1-6.5 \times 10^{10}~\rm{K\, km \,s}^{-1}\,\rm{pc}^2$, and the mass-to-light conversion factor $\alpha_{\rm{CO(1-0)}} = 5.2-17~\rm{M}_\odot\, \rm{pc}^{-2} (\rm{K\, km \,s}^{-1})^{-1}$ corresponding to the 16-84th percentile ranges.
The spatially unresolved measurements, combined with a fixed GDMR, and [CO]/[H$_2$] values used in this initial modelling, 
prevent a more tightly constrained molecular gas mass. Nevertheless, the derived conversion factor is 
significantly higher than the commonly assumed ULIRG value of 
$0.8~\rm{M}_\odot\, \rm{pc}^{-2} (\rm{K\, km \,s}^{-1})^{-1}$ \citep{DownesSolomon1998}, despite the total IR luminosity of J154506 exceeding 10$^{12} \rm{L}_{\odot}$.\\

Despite providing only a global view of J154506's molecular gas and dust properties, our analysis yields reliable ISM properties. This confidence stems from the extensive coverage of the source's dust emission and thorough sampling of its CO ladder. These data, combined with the narrowness of the marginalised posterior distributions of the parameters (see Appendix~\ref{App:Dust}), reinforce the robustness of this initial characterisation, even if a more comprehensive analysis remains beyond the scope of the present study.
While future spatially resolved studies of the dust and cold gas will allow for more precise estimates, 
our initial findings indicate that this submm-selected lensed object exhibits less extreme molecular gas excitation conditions compared to other known high-redshift DSFG and QSO systems. 
Moreover, its substantial molecular gas reservoir suggests the potential to evolve into a hyperluminous infrared galaxy (HyLIRG) at cosmic noon ($z=2-3$) upon reaching the peak of its starburst activity.

% %%%%%%%%%%%%%%%%%%%
\subsection{[CII] as a tracer of cold ISM}
%%%%%%%%%%%%%%%%%%%%%%%%%%%%%%%%%%%%%%%%%%%%%%%%%%%
\label{Sect:CIILineProperties}
 
The [CII] 158 $\mu$m line is the brightest FIR line and the dominant coolant of neutral and ionised gas. 
It originates primarily in photo-dissociation regions (PDRs), 
where far-ultraviolet photons from young massive stars interact with molecular clouds, 
but it can also arise from the diffuse ionised medium and cold neutral medium \citep[e.g.][]{Lagache2018}. 
Since our observations are spatially unresolved and have a modest S/N ($<5$), 
our analysis derives global constraints 
based on the ability of [CII] to probe gas content and trace SF activity \citep[i.e.][]{CarilliWalter2013} 
up to very high redshift \citep[e.g. ][]{Capak2015, Bouwens2022}.

%%%%%%%%%%%%%%%%%%%
We derive an observed [CII] luminosity of $3.2\times10^{10}\rm L _{\odot}$ (L$^{\prime}_{\rm{[CII]}}=14.9\times10^{10} 
~\rm{K}~\rm{km}~\rm{s}^{-1}~\rm{pc}^{2}$), corresponding to an intrinsic L$_{\rm{[CII]}}$ of $5.4\times10^{9}\rm L _{\odot}$.
This corresponds to an L$_{\rm{[CII]}}$ to L$_{\rm{FIR}}$ ratio of 8.3 to $9.9
\times10^{-4}$, 
depending on the L$_{\rm{FIR}}$ adopted
(see Sections~\ref{Sect:DustProperties} and \ref{Sect:GasProperties}).
This ratio is comparable to values (L$_{\rm{[CII]}}$/L$_{\rm{FIR}}= 3.6- 25.7\times10^{-4}$)
reported by \cite{Gullberg2015} for 20 SPT galaxies at $z=2.1-5.7$ 
with [CII] luminosities of $1.4-9.2\times 10^{10}\rm{L}_{\odot}$.

%%%%%%%%%%%%%%%%%%%
Using the high-redshift relations from \cite{deLooze2014}, 
we derive an SFR of 840~M$_{\odot}\rm{yr}^{-1}$. 
This [CII]-derived SFR is remarkably consistent with that obtained 
from the median total infrared luminosity (L$_{\rm TIR}\sim 6 \times 10^{12}\rm{L}_{\odot}$). 
This finding is noteworthy, as studies of infrared-luminous galaxies (L$_{\rm{IR}}>10^{12}$) galaxies 
commonly report a suppression of [CII] emission relative to infrared emission, 
known as the `[CII] deficit' \citep[e.g. ][]{DiazSantos2013, Lutz2016, HerreraCamus2018}.
For J154506, this suggests that the [CII] emission serves as an efficient cooling channel, suggesting that physical conditions (e.g. radiation field intensity and gas density) 
do not substantially suppress [CII] emission. This behaviour aligns with observations of 
other high-redshift sources, where [CII] effectively traces SF
\citep[e.g. ][]{Carniani2020, Schaerer2020}.

% %%%%%%%%%%%%%%%%%%%
\subsection{Possible AGN contribution}
% %%%%%%%%%%%%%%%%%%%
The intrinsic IR-based SFR of J154506 is remarkably high, estimated at $\gtrsim $840 M$_{\odot}\rm{yr}^{-1}$,  
approaching the limits of typical maximal starburst systems \citep[i.e.][]{Casey2014, Bethermin2015, Tacconi2020}.
This raises the possibility that the total infrared luminosity, and consequently the derived SFR, 
might be contaminated by emission from a dust-obscured AGN.
The AGN fraction ($f_{\rm AGN}$) can, however, be constrained using the upper limit on the 
Multiband Imaging Photometer for Spitzer (MIPS) 24\,$\mu$m flux, 
which probes the rest-frame $\sim5\,\mu$m emission at $z=3.75$. 
For this analysis, we assume that foreground extinction from the Lupus-I molecular cloud at 24\,$\mu$m is negligible. 
For J154506, the ratio $\nu \rm L_\nu(5\ \mu\rm m) / \rm L_{\rm IR} < 0.04$, 
suggesting that the source is SF-dominated, as this value is well below the threshold of 0.1 
typically indicative of a significant AGN contribution \citep{HernanCaballero2009}. 

This aligns with the near-unity T$_{\rm{k}}$ / T$_{\rm{d}} \approx 1.1$, suggesting 
that ISM heating is largely dominated by star formation rather than AGN activity. 
In AGN-dominated systems, strong X-ray and UV radiation from the central source 
influence the ISM through cosmic ray heating, ionisation, and turbulence. This can significantly enhance T$_{\rm{k}}$ while leaving the dust temperature relatively low, 
leading to T$_{\rm{k}}$/T$_{\rm{d}} \gg 1$ \citep[e.g.][]{Schleicher2010, Papadopoulos2010, Narayanan2014}. 
Observational studies confirm that AGN hosts and starburst-dominated HyLIRGs typically exhibit T$_{\rm{k}}$/T$_{\rm{d}} > 2$ \citep[e.g. ][]{Harrington2021}, 
whereas moderate star-forming systems tend to maintain lower ratios of T$_{\rm{k}}$/T$_{\rm{d}} = 1-1.5$
\citep[e.g.][]{Daddi2015, Valentino2020}.
In addition, the CO line SED indicates that the molecular gas excitation in this system 
is not driven by AGN-related X-ray heating or intense mechanical shocks, 
which typically produce enhanced high-J CO emission. However, we acknowledge that a minor AGN contribution cannot be definitively ruled out. Deeper observations are required to more stringently constrain the high-J emission. 
Current samples with well-mapped CO SLEDs 
remain heavily biased towards low-redshift galaxies, lower-J transitions, and extreme systems, 
limiting direct comparisons with high-redshift DSFGs.

%%%%%%%%%%%%%%%%%%%%%%%%%%%%%%%%%%%%%%%%%%%%%%%%%%%
\section{Summary and conclusions}
%%%%%%%%%%%%%%%%%%%%%%%%%%%%%%%%%%%%%%%%%%%%%%%%%%%
We confirm the extragalactic and strongly lensed nature of J154506, 
an extremely submm-bright and enigmatic source that was initially thought to be a (sub)stellar object 
due to its location towards the Lupus-I molecular cloud. 
By combining archival data with new submm spectral scans, 
we spectroscopically confirm its redshift and characterise the gas, 
dust, and physical properties of J154506, yielding the following key results:

\begin{itemize}
    \item 
    Using ALMA/ACA and LMT/RSR, we detect two significant (S/N>5) broad ($\sim600~\rm{km\,s}^{-1}$) emission lines at 97.0 and 145.5~GHz, 
    corresponding to CO(4-3) and CO(6-5), respectively.
    These detections yield a spectroscopic redshift of z$_{\rm{spec}}=3.7515\pm 0.0005$ and
    confirm J154506 as an extragalactic source, consistent with the arc-shaped structure in ALMA continuum images. 

    \item
    We also detect several lower S/N ($=2-3$) lines:  
    [CI], HCN(4-3), and two H$_2$O$^+$ transitions (H$_2\rm{O}^+(2_{02}-1_{11})_{(3/2-1/2)}$ and H$_2\rm{O}^+(2_{11}-2_{02})_{(3/2-3/2)}$), along with additional tentative detections.

    \item 
    We performed gravitational lens modelling using \textsc{PyAutoLens}, 
    fitting the ALMA B6 and B7 continuum data directly in the uv-plane 
    with a singular isothermal ellipsoid (SIE) lens model and a Sérsic source profile, 
    yielding an average magnification factor of $\mu = 6.0 \pm 0.4$. 
    Although observing conditions prevented a spectroscopic redshift confirmation 
    of the foreground lens with MUSE and FORS2, these optical observations were crucial 
    for accurately pinpointing its position. 

    \item 
    We detect the [CII] 158~$\mu$m fine-structure line  
    at 400~GHz using APEX/nFLASH in three independent spectra.
    From the median spectrum, we derive an intrinsic luminosity of $5.4\times10^{9}\rm L_{\odot}$ 
    after correcting for magnification, and an SFR of $\sim 840\rm{M}_{\odot}\rm{yr}^{-1}$.

    \item 
    The simple modified blackbody fit to the FIR photometry of J154506 
   yields a relatively cold dust temperature (30-48~K), an emissivity index $\beta$ of $2.0$, and a dust mass of $4.5\times10^{9}~\rm{M}_{\odot}$.

    \item 
    The combined modelling of the dust SED and CO excitation ladder constrains the dust temperature to 38~K and suggests that 
    the FIR emission arises primarily from moderately dense rather than 
    compact, high-pressure environments typical of extreme starbursts or AGNs. 
    Additionally, we find that the CO excitation ladder peaks near CO(5-4)
    and is dominated by a slightly denser molecular gas. 
    Its sharp turnover further supports the interpretation that J154506 is a highly star-forming galaxy, 
    but not a QSO or an extreme system.
    The derived near-unity kinetic-to-dust temperature ratio suggests a minor AGN contribution for J154506.
    
\end{itemize}

Our results highlight the importance of mapping even low Galactic latitudes 
when searching for such extreme and scarce sources and demonstrate their potential to 
probe the ISM properties of high-redshift galaxies in unprecedented detail. 
A spectroscopic confirmation of the foreground is required  
to further characterise this extreme system.
%--------------------------------------------------------------------

%--------------------------------------------------------------------
\begin{acknowledgements} 
We thank the anonymous referee for a thoughtful and detailed review, which has significantly improved the clarity and quality of this manuscript.
This paper makes use of the following ALMA data: ADS/JAO.ALMA\#2015.1.00512.S; ADS/JAO.ALMA\#2017.1.00303.S; ADS/JAO.ALMA\#2018.1.00126.S; ADS/JAO.ALMA\#2019.1.00245.S;  ADS/JAO.ALMA\#2021.2.00097.S; ADS/JAO.ALMA\#2023.1.00251.S. 
ALMA is a partnership of ESO (representing its member states), NSF (USA) and NINS (Japan), together with NRC (Canada), MOST and ASIAA (Taiwan), and KASI (Republic of Korea), in cooperation with the Republic of Chile. The Joint ALMA Observatory is operated by ESO, AUI/NRAO, and NAOJ.
M.A. is supported by FONDECYT grant number 1252054, and gratefully acknowledges support from ANID Basal Project FB210003 and ANID MILENIO NCN2024\_112. 
A.S.M. acknowledges support from ANID/Fondo 2022 ALMA/31220025. 
E.F.-J.A. acknowledges support from UNAM-PAPIIT projects IA102023 and IA104725, and from CONAHCyT Ciencia de Frontera project ID: CF-2023-I-506.
MIR acknowledges the support of the Spanish Ministry of Science, Innovation and Universities through the project PID-2021-122544NB-C43. 
M.S. was financially supported by Becas-ANID scholarship \#21221511, and also acknowledges support from ANID BASAL project FB210003. 
This work would not have been possible without the long-term
financial support from the Mexican Humanities, Science and Technology
Funding Agency, CONAHCYT (Consejo Nacional de Humanidades, Ciencias y
Tecnologías), and the US National Science Foundation (NSF), as well as
the Instituto Nacional de Astrofísica, Óptica y Electrónica (INAOE)
and the University of Massachusetts, Amherst (UMass). The operation of
the LMT is currently funded by CONAHCYT grant \#297324 and NSF grant \#2034318. The data described in this paper include LMT observations conducted
under the scientific program 2024-S1-MX-11. The authors acknowledge the scientific and technical support
of the LMT staff during the observations and generation of data
products provided to the authors.

\textit{Software:}
In addition to the software mentioned in the main text, this work also employed Astropy, a community-developed core Python package
for Astronomy \citep{AstropyCollaboration2013, astropy2022}; Matplotlib \citep{Hunter2007}; Numpy; SciPy \citep{SciPy2020}; Photutils \citep{Bradley2016}; Interferopy \citep{Boogaard2021}; Astroquery \citep{astroquery2019}, and mpdaf \citep{mpdaf2016}.
\end{acknowledgements}
%-------------------------------------------------------------------

%--------------------------------------------------------------------
%BIBLIOGRAPHY
\bibliographystyle{aa} % style aa.bst
\bibliography{all} % your references Yourfile.bib

%-------------------------------------------------------------------
\begin{appendix}

\onecolumn

%%%%%%%%%%%%%%%%%%%%%%%%%%%%%%%%%%%%%%%%%%%%%%%%%%%%%%%%%%%%%%%%%%%%%
\section{ALMA archival data}
%%%%%%%%%%%%%%%%%%%%%%%%%%%%%%%%%%%%%%%%%%%%%%%%%%%

\label{App:ALMAarchival}

\begin{figure*}[!h]
\includegraphics[width=0.252\textwidth]{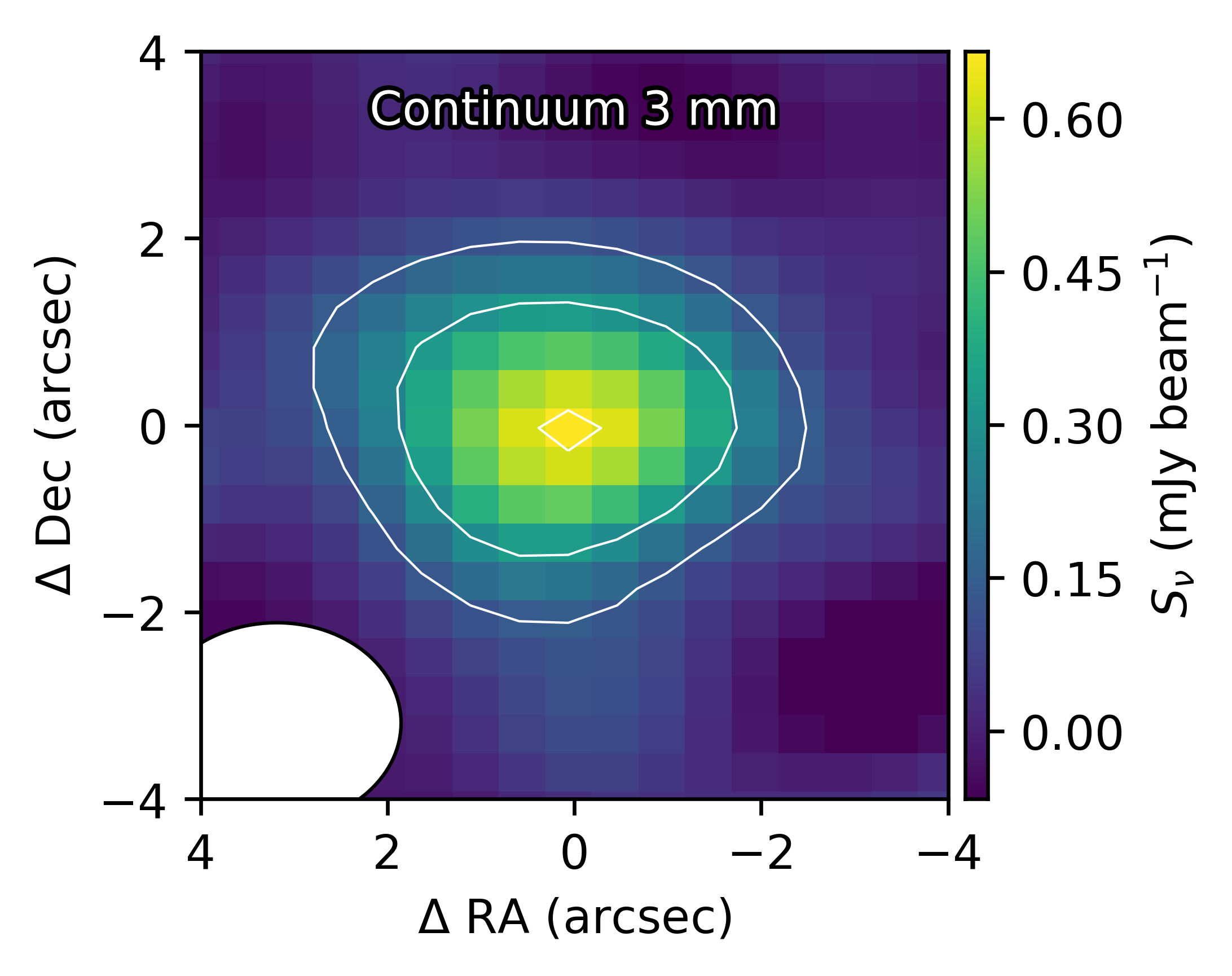}
\includegraphics[width=0.24\textwidth]{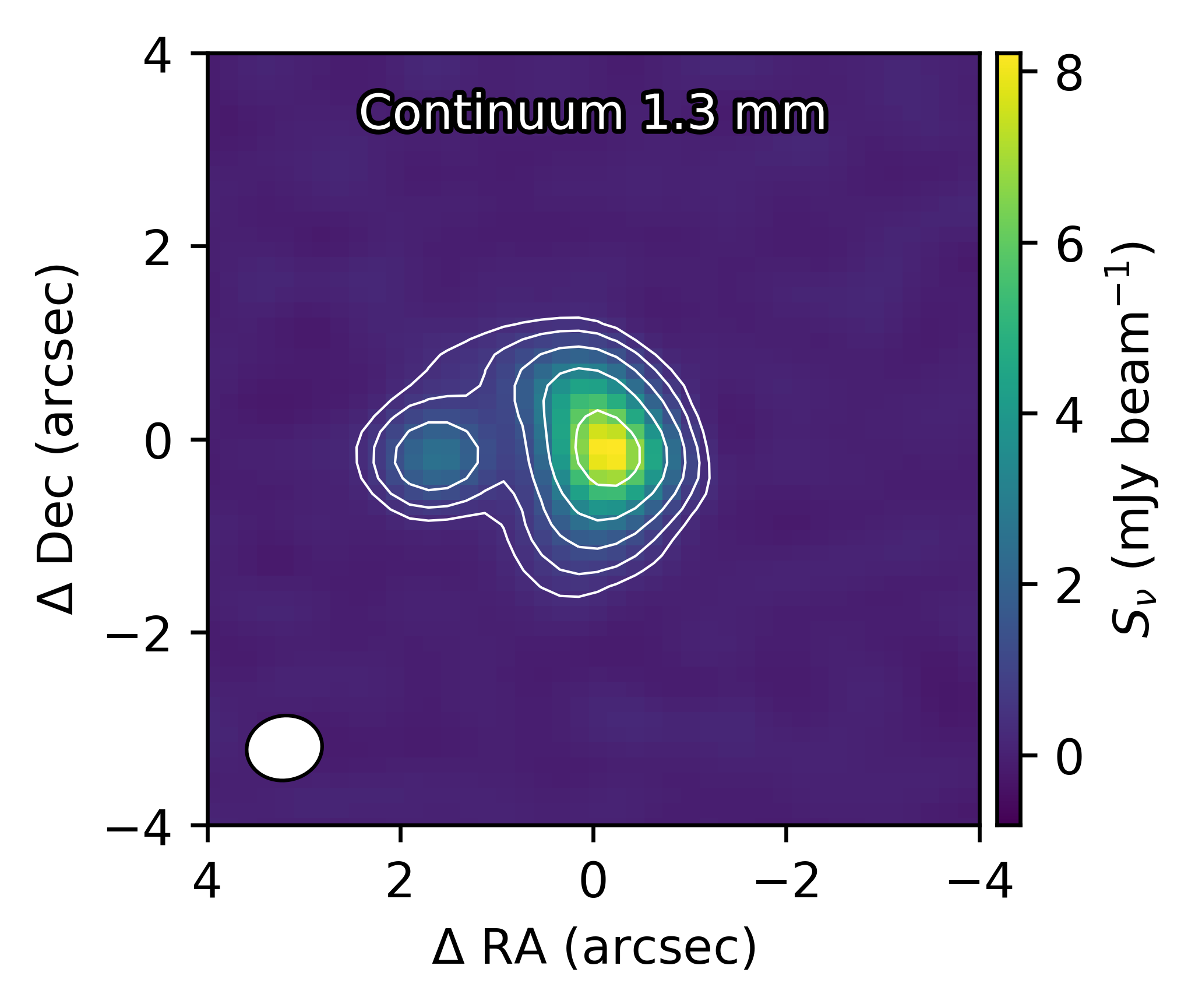}
\includegraphics[width=0.24\textwidth]{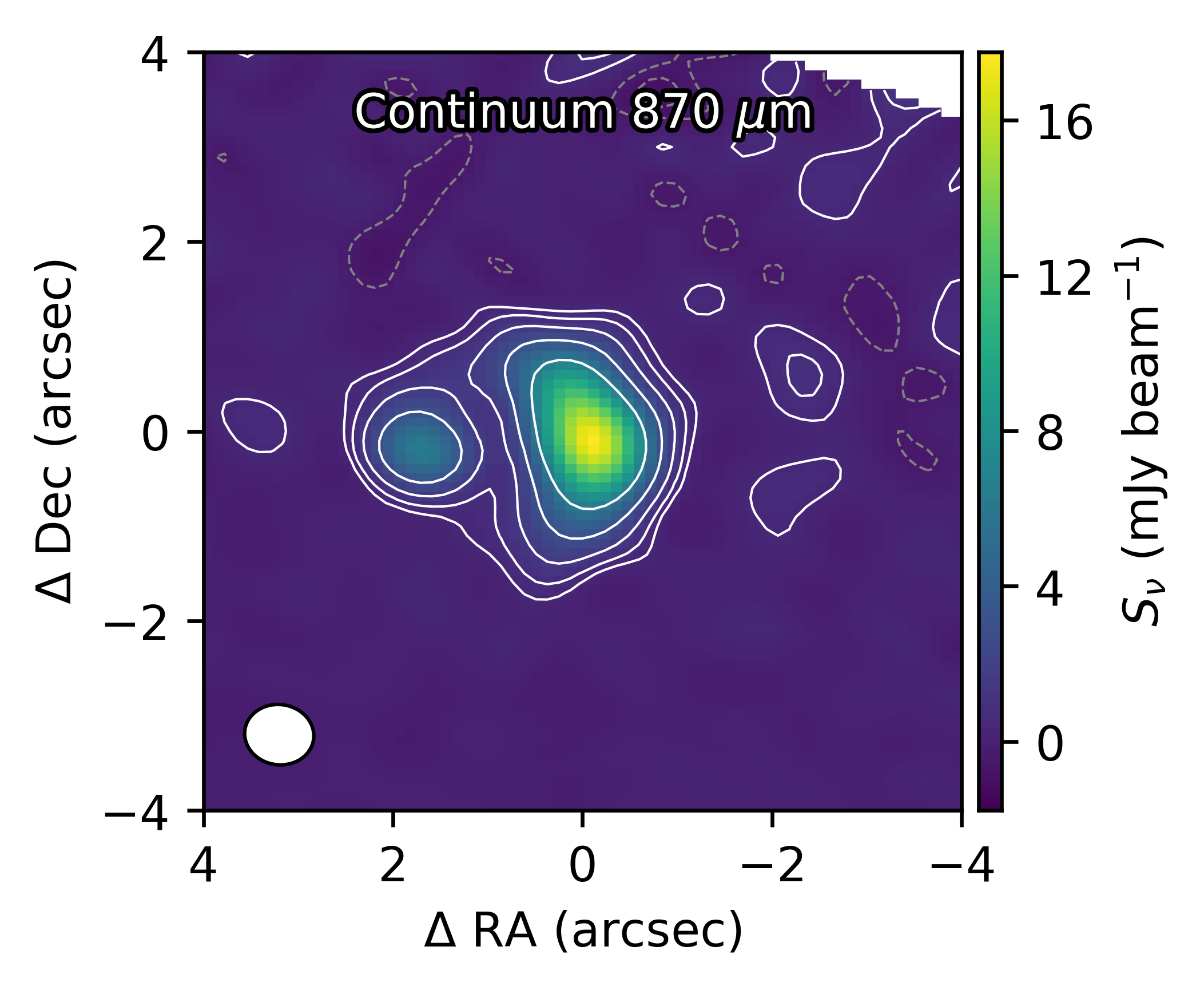}
\includegraphics[width=0.252\textwidth]{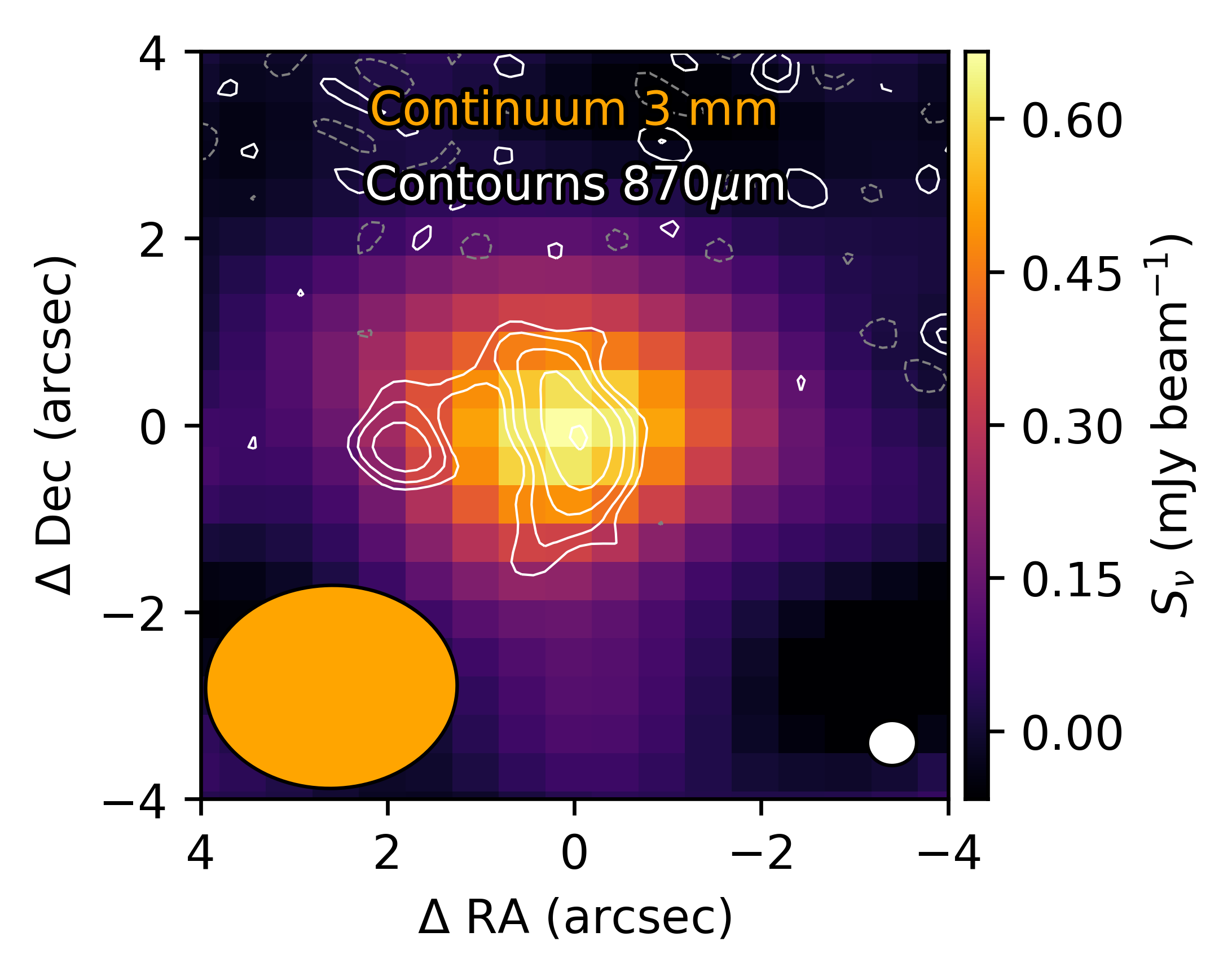}
\caption{\textit{Left panels: }4"$\times$4" ALMA Bands 3, 6, and 7 imaged with \textsc{Briggs} weighting (robust=0.5) centred on the gravitationally lensed galaxy J154506 are shown from left to right. -4, -2, 2, 4, 8, 16, and 32 $\sigma$ contours are shown with solid and dashed white lines for positive and negative values, respectively. {Right panel: } Band 7 contours (\textsc{Briggs} weighting, robust=1.0) on top of the Band 3 data, for comparison purposes.}
\label{Fig:ALMAarchivalData}
\end{figure*}

The specifics of the ALMA archival data in the Lupus-I molecular cloud that allowed us to confirm
the extragalactic nature of the background source in J154506 are presented in this appendix.
Fig.~\ref{Fig:ALMAarchivalData} showcases the unresolved and resolved emission in ALMA Bands 3, 6, and 7. 
Table~\ref{table:ALMAdata} provides details of the specific projects, the central sky frequencies, sensitivity, and resolution.
Details on data reduction are provided in Section~\ref{Sect:ALMA ancillary data}.
%-------------------------------------- 
%-------

%--------------------------------------------------------------------
\renewcommand{\arraystretch}{1.3}
%--------------------------------------------------------
\begin{table*}[!h]
\caption{Overview of ALMA archival continuum observations for J154506.}
\label{table:ALMAdata}      
\centering          
\begin{tabular}{ l c c c c c }     
\hline
            & Cycle 3 B6    & Cycle 5 B3$^{b}$ & Cycle 6 B7 & Cycle 7 B3 & Cycle 8 B6 \\
\hline   
Project ID  & 2015.1.00512.S & gray2017.1.00303.S & 2018.1.00126.S & 2019.1.00245.S & 2021.2.00097.S \\
Array & 12m & 12m & 12m & 12m & 7m \\
Frequency [GHz] &216.1-234.4  & 93.1-107.3 & 333.1-348.7 & 93.1-107.3& 216.0-234.5\\
Beam [arcsec$^{2}$]$^{a}$ &  \ang{;;0.79} $\times$ \ang{;;0.67} &  ---  & \ang{;;0.88} $\times$ \ang{;;0.75} &  \ang{;;2.69} $\times$ \ang{;;2.17} &  \ang{;;6.47} $\times$ \ang{;;4.22}\\
Spatial resolution [arcsec] & 0.75 & 2.18 & 0.75 & 2.51 & 4.79 \\
Sensitivity [mJy beam $^{-1}$] & 0.05 & 0.05 & 0.12 & 0.05 & 0.91 \\
\hline
\end{tabular}\\
$^{a}$ The beam size corresponds to the imaging with \textsc{Briggs} weighting (robust=0.5).\\
$^{b}$ In this work, we do not analyse Cycle 5 B3 semipass data, but details are shown in the table for completeness.\\
\end{table*}
%--------------------------------------------------------

%%%%%%%%%%%%%%%%%%%%%%%%%%%%%%%%%%%%%%%%%%%%%%%%%%%%%%%%%%%%%%%%%%%%%
\section{LMT spectral features }
%%%%%%%%%%%%%%%%%%%%%%%%%%%%%%%%%%%%%%%%%%%%%%%%%%%
\label{App:lmt}

This appendix provides additional LMT/RSR zoomed spectra, supplementing the analysis presented in the main text and Figure \ref{Fig:LMTspectra}.

\begin{figure}[!h]
 \includegraphics[width=1.\textwidth]{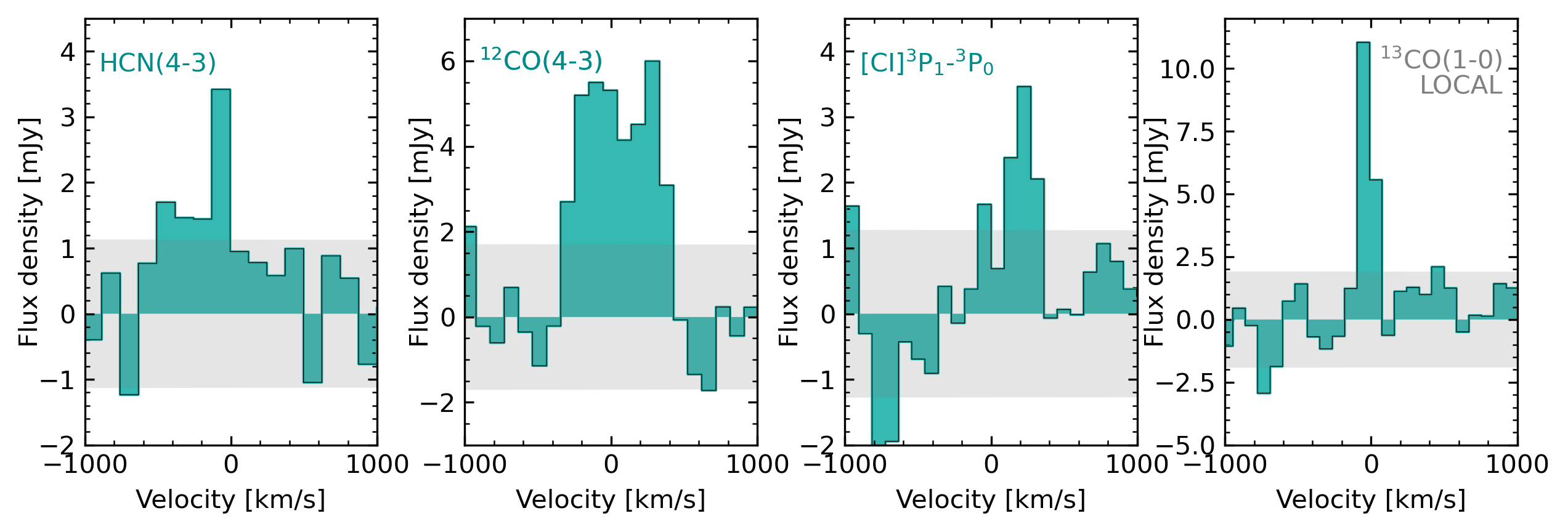}
 \caption{
 LMT/RSR spectra at their native (unbinned) resolution at the position of the detected lines at $z=3.5715$ (\textit{three first panels; }HCN(4-3), $^{12}$CO(4-3), and [CI]$^{3}$P$_{1}$-$^{3}$P$_{0}$ from left to right) and the Lupus molecular cloud (\textit{right panel; }local $^{13}$CO(1-0)). Note that the peak flux density is slightly higher than for the binned spectrum shown in the bottom panel of Figure~\ref{Fig:LMTspectra}.
}
\label{Fig:LMTspectra_pannels}
\end{figure}

\clearpage

%%%%%%%%%%%%%%%%%%%%%%%%%%%%%%%%%%%%%%%%%%%%%%%%%%%%%%%%%%%%%%%%%%%%%
\section{ACA spectral scans }
%%%%%%%%%%%%%%%%%%%%%%%%%%%%%%%%%%%%%%%%%%%%%%%%%%%
\label{App:aca}

\begin{figure}[!h]
\includegraphics[width=1.\textwidth]{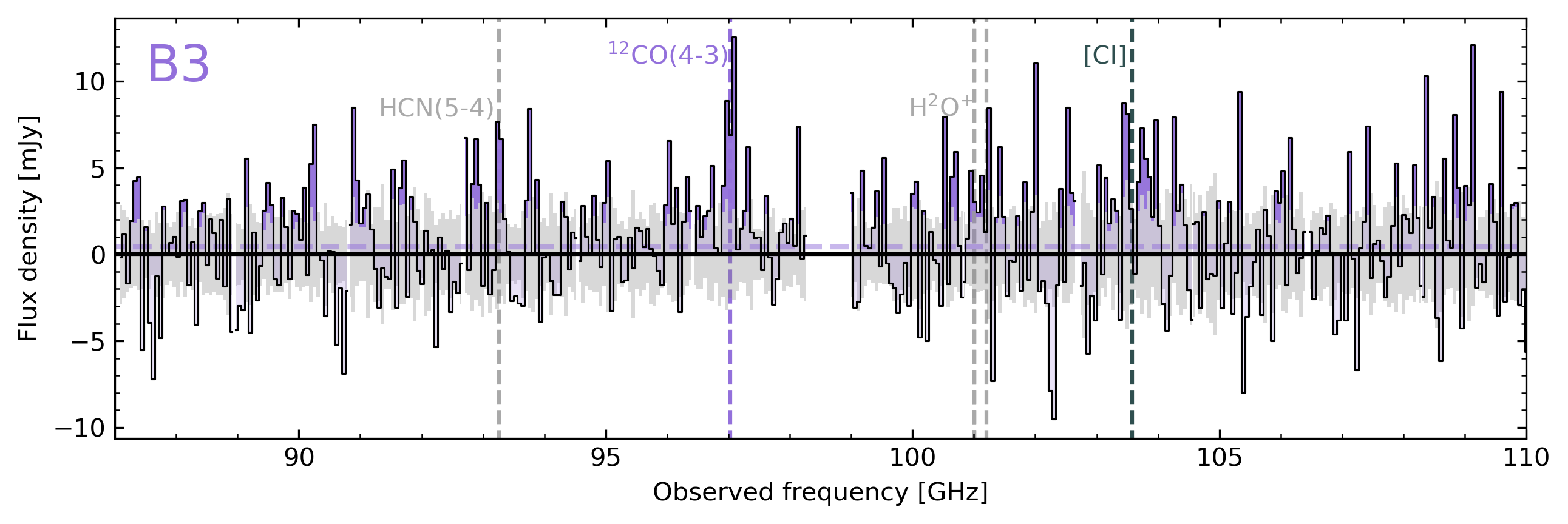}
\includegraphics[width=1.\textwidth]{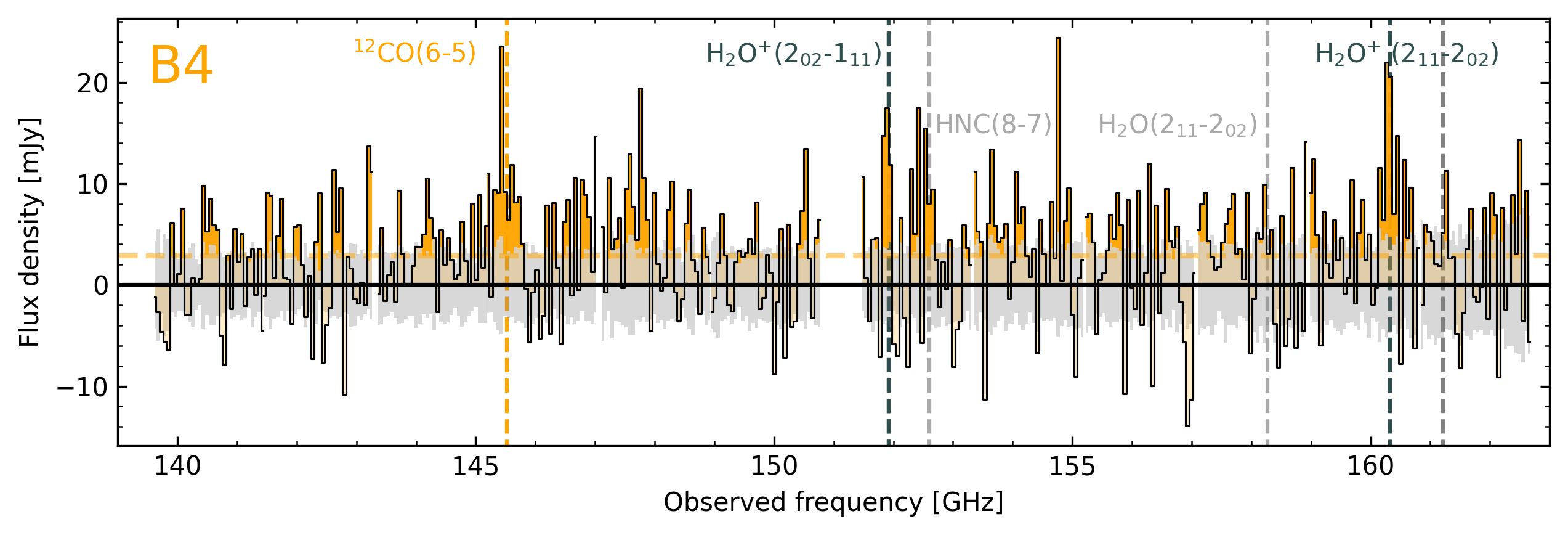}
 \caption{
 \textit{Top panels}: 
 ALMA/ACA B3 (\textit{first row}, purple) and B3 (\textit{second row}, orange) spectrum of J154506 (colour) and associated errors (grey) before continuum subtraction. The transitions above 
 S/N=5 and 3 compatible with J154506’s redshift ($z=3.7515$) are marked with vertical dashed lines, and labelled in colour, and dark grey, respectively. Other S/N$>$1 consistent tentative transitions are highlighted in light grey. The medium continuum emission is highlighted with a coloured horizontal dashed line.
 \textit{Bottom panels}: Continuum subtracted spectra at the position of the S/N$>$3 detected lines at z$=3.5715$ in B3 (\textit{left panels; }$^{12}$CO(4-3), and [CI]$^{3}$P$_{1}$-$^{3}$P$_{0}$) and B4 (\textit{right panels; }$^{12}$CO(4-3), and H$_2\rm{O}^+(2_{11}$-$2_{02})_{(3/2-3/2)}$.
}
\label{Fig:ACAspectra}
\end{figure}

This appendix presents the spectral scans obtained with the Atacama Compact Array (ACA).  Figure \ref{Fig:ACAspectra} displays the observed spectra in Band 3 (B3) and Band 4 (B4), along with an analysis of potential line identifications.

\clearpage

%--------------------------------------------------------
\setlength{\tabcolsep}{5pt}

\begin{table*}[!h]
\caption{Overview of ALMA/ACA spectral observations of J154506}   
\centering          
\begin{tabular}{ l c c c c c c}     
\hline \hline
Project ID  & \multicolumn{6}{c}{2023.1.00251.S (Cycle 10, ACA)}  \\
\hline   
Scan    & B3a   & B3b  & B3c  & B4a  & B4b  & B4c   \\
Frequency, LSB [GHz]& 87.0-90.8&90.8-94.5& 94.5-98.3& 139.6-143.3&143.3-147.1& 147.1-150.8 \\
Frequency, USB [GHz]& 98.9-102.7 &102.7-106.4& 106.4-110.2& 151.5-155.2&155.2-158.9& 158.9-162.7 \\
Beam [arcsec$^{2}$] & \ang{;;18.9} $\times$ \ang{;;8.9}& \ang{;;18.6} $\times$ \ang{;;8.7} & \ang{;;20.0} $\times$ \ang{;;8.4} & \ang{;;11.0} $\times$ \ang{;;6.5} & \ang{;;12.1} $\times$ \ang{;;5.7} & \ang{;;11.3} $\times$ \ang{;;5.3}\\
Spatial resolution [arcsec]  & 11.22 & 10.90& 12.07& 7.46 & 7.01 & 5.64\\
Sensitivity [mJy beam $^{-1}$] & 0.23 & 0.26 & 0.23 & 0.34& 0.33 & 0.37 \\
Spectral resolution [km s$^{-1}$]$^{a}$ & 12 & 3 & 3 & 2& 2 & 2\\
\hline
\hline   
\label{table:ACAdata}   
\end{tabular}
$^{a}$Note that this is the native spectral resolution, but spectral windows were binned to $\approx100$~ kms$^{-1}$ to increase sensitivity as detailed in Section~\ref{Sect:ACAobservations}.
\end{table*}
%--------------------------------------------------------

% \clearpage

%%%%%%%%%%%%%%%%%%%%%%%%%%%%%%%%%%%%%%%%%%%%%%%%%%%%%%%%%%%%%%%%%%%%%
\section{Photometry}
%%%%%%%%%%%%%%%%%%%%%%%%%%%%%%%%%%%%%%%%%%%%%%%%%%%
\label{App:Photometry}

This Appendix contains the archival and new photometric measurements used throughout this work.

%--------------------------------------------------------
\begin{table}[!h]
\caption{Photometry (in mJy) used in this work to derive the physical properties of J154506.
}             
\label{table:Photometry_J1545}      
\centering          
\begin{tabular}{ l c c c l c c c}     
\hline\hline       
Telescope/ & Band & Flux density & Ref.      &   Telescope/ & Band & Flux density & Ref.           \\               
instrument   & ($\mu$m)  & (mJy)         &   &   instrument   & ($\mu$m)  & (mJy)                   \\               
\hline   
MUSE        & 0.48-0.93 & (0.6 $\pm$ 0.2) $\times 10^{-3}$&  This work & SMA & 1300  & 20.8 $\pm$ 1.9   & (1)            \\    
MIPS        & 24        & <0.3               & (1)$^{a}$       &    ACA B4c  &   1940      & 4.9 $\pm$ 1.4    & This work             \\    
SPIRE       & 250       & 40.9 $\pm$ 12.7  & (1)         &    ACA B4b  &   1980      & 4.8 $\pm$ 1.5    & This work             \\     
SPIRE       & 350       & 109.4 $\pm$ 11.4 & (1)         &    ACA B4a  &   2030      & 3.8 $\pm$ 1.2    & This work             \\     
SPIRE       & 500       & 134.9 $\pm$ 11.9 & (1)         &    ACA B3c  &   2930      & 1.1 $\pm$ 0.4    & This work             \\     
SMA         & 890       & 69.7 $\pm$ 12.1  & (1)         &    ALMA B3  &   3000      & 0.9 $\pm$ 0.2    & This work            \\   
ALMA B7     & 890       & 46.5 $\pm$ 3.7   &  This work  &    ACA B3b  &   3040      & 0.8 $\pm$ 0.3    & This work             \\           
AzTEC       & 1100      & 43.9 $\pm$ 5.6   & (1)         &    ACA B3a  &   3160      & 0.3 $\pm$ 0.2    & This work             \\     
ALMA B6     & 1300      & 21.3 $\pm$ 1.8   &  This work  &    ATCA     &   7000      & 210 $\pm$ 35 $\times 10^{-3}$ & (1)       \\ 
ACA B6      & 1300      & 18.0 $\pm$ 4.4   &  This work  &    VLA      &   60000     & 66 $\pm$ 5 $\times 10^{-3}$  & (1)       \\ 
\hline \hline                
\end{tabular}\\
$^{a}$ For archival multi-wavelength data we refer the reader to the details provided in \citealt{Tamura2015} (1).\\
\end{table}
%--------------------------------------------------------

%%%%%%%%%%%%%%%%%%%%%%%%%%%%%%%%%%%%%%%%%%%%%%%%%%%%%%%%%%%%%%%%%%%%%
\section{Simultaneous modelling of the observed continuum and emission lines}
%%%%%%%%%%%%%%%%%%%%%%%%%%%%%%%%%%%%%%%%%%%%%%%%%%%
\label{App:CO_SLED}

The \tuner\ model solves for the non-LTE radiative transfer of the lines in the LVG approximation and effectively computes the line brightness temperatures by simultaneously fitting the dust continuum and line emission. The dust temperature and continuum emission serves as an additional temperature floor on top of the blackbody CMB radiation at the redshift of the object. We additionally constrain the parameter space by implementing a physically motivated range of the T$_{\rm kin}$/T$_{\rm dust}$, and couple the \Htwo\ density and gas kinetic temperature with a power-law slope index. Still, there can be a wide-ranging and highly degenerate parameter space.  We have applied the Markov chain Monte Carlo \textit{emcee} Python package \citep{ForemanMackey2013} using 100 walkers and 50 autocorrelation times and uniform priors.

Here we have allowed the following parameters to be optimized and show their 1D and 2D marginalized posterior distributions: dust temperature and dust emissivity index T$_{\rm dust}$ (K) and $\beta_{\rm dust}$, gas kinetic temperature T$_{\rm kin}$ (K), turbulent velocity dispersion $\Delta  V$ (km\,s$^{-1}$), emitting size radius $r$ (pc), log(n(H$_2$))[cm$^{-3}$], and gas kinetic temperature to density power law slope ($\beta_{\rm T}$) which couples the gas kinetic temperature to the gas volume density by a power-law index, $\beta_{\text{T}_{\text{kin}}}$, as $T_{\text{kin}} \propto \log(n(\text{H}_2))^{\beta_{\text{T}_{\text{kin}}}}$, such that the more diffuse gas tends to have higher gas kinetic temperatures. The magnification of the object (see Section~\ref{Sect:Lens_model_magnification}) is fixed in the model to therefore provide the intrinsic source properties \citep[see][]{Weiss2007}. 

\renewcommand{\arraystretch}{1.3}
%--------------------------------------------------------
\begin{table*}[!h]
\caption{Parameter space explored by our \textit{Turbulence} model}
\label{table:TurbModel}      
\centering  
\begin{tabular}{ c c c c c c c  }     
\hline
\hline
T$_{\rm kin}$/T$_{\rm dust}$ & $\Delta  V$ [km\,s$^{-1}$]  & $\gamma_{\rm T}$ & $\beta_{\rm dust}$  & r[kpc] &  T$_{\rm kin}$[K]& log(n(H$_2$))[cm$^{-3}$]\\
\hline
0.5---6 & 1---150  & -0.3---0.005 & 1.5---2.8   & 0.1---8000 &  T$_{\rm{CMB}}$$^{a}$---600   &  1--- 7 \\ 
\hline
\hline
\end{tabular}

$^{a}$The minimum temperature is the temperature of the CMB at z=3.75 plus a temperature floor of 10K.
\end{table*}
%--------------------------------------------------------

\begin{figure*}[!h]
    \centering
    \includegraphics[width=1.\textwidth]{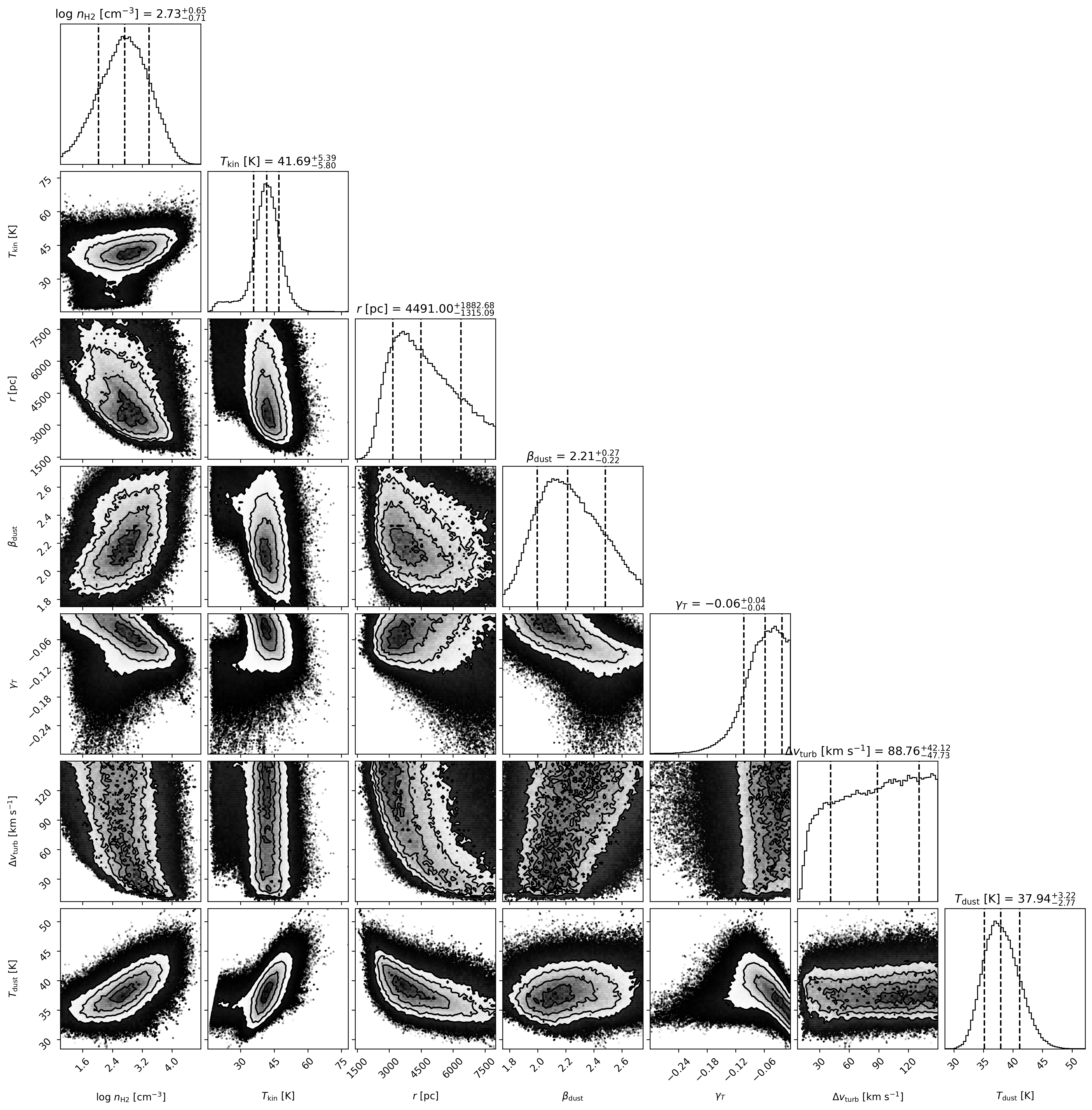}
    \caption{
    Derived marginalized 1D and 2D posterior distributions for H$_2$ gas density log(n(H$_2$)), gas kinetic temperature (T$_{\rm kin}$), emitting radius (r), dust emissivity index ($\beta_{\rm dust}$), density-to-gas kinetic temperature power-law slope index ($\gamma_{\rm T}$), turbulent velocity dispersion ($\Delta  V$), and dust temperature (T$_{\rm dust}$) are shown. Note that the gas density is in log10 units. The log([CO]/[\Htwo]) and GDMR are fixed to -4.0 and 150, respectively.
    }
\end{figure*}

\clearpage

%%%%%%%%%%%%%%%%%%%%%%%%%%%%%%%%%%%%%%%%%%%%%%%%%%%%%%%%%%%%%%%%%%%%%
\section{MERCURIUS FIR FIT}
%%%%%%%%%%%%%%%%%%%%%%%%%%%%%%%%%%%%%%%%%%%%%%%%%%%
\label{App:Dust}

This appendix provides the posterior distribution of the main parameters considered by \textsc{MERCURIUS} during the FIR fitting and dust properties estimation process for both the self-consistent and the entirely optically thin scenario.

\begin{figure*}[!h]
    \includegraphics[width=1.\textwidth]{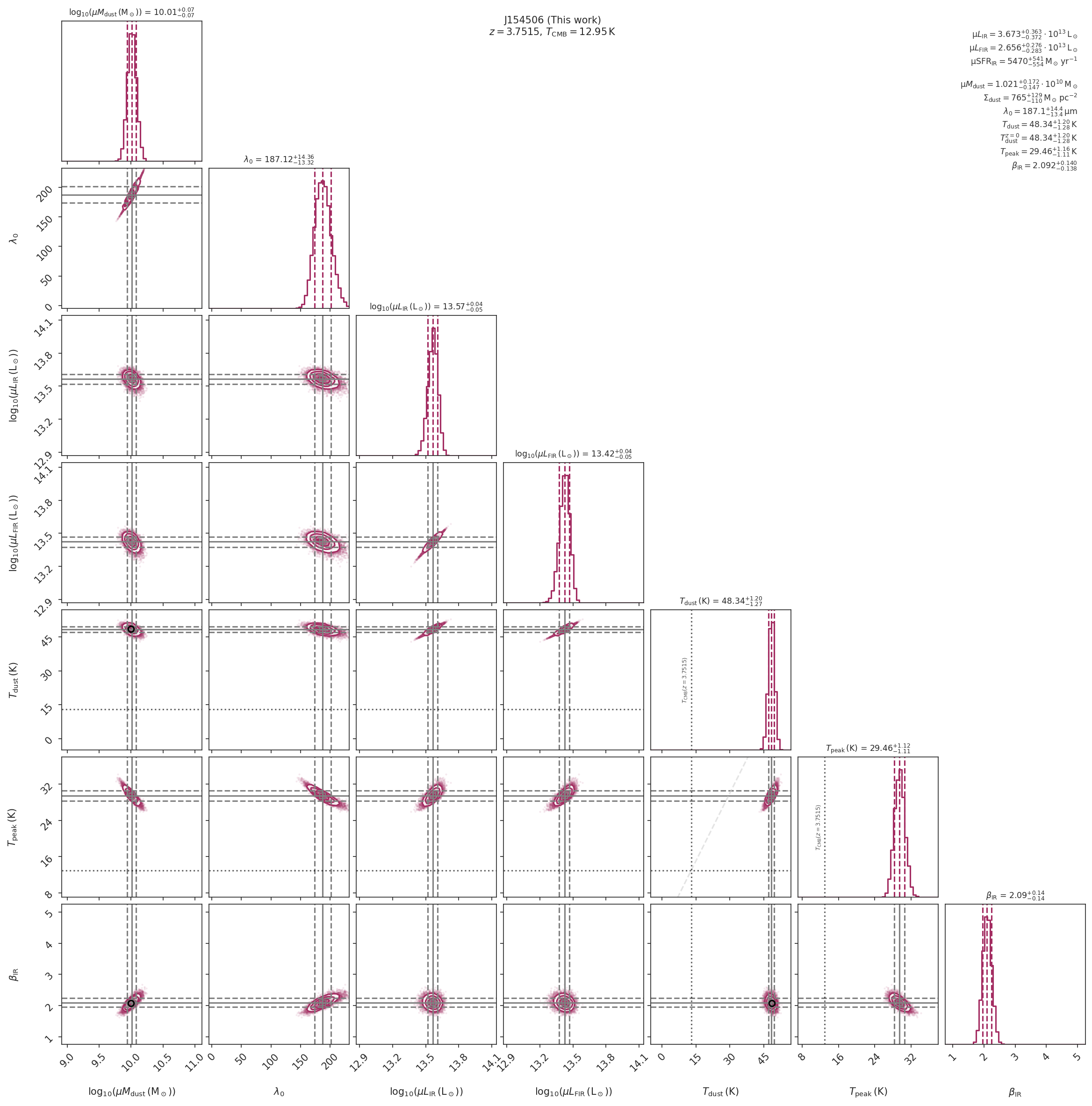}
    \caption{
    Corner plot of the posterior distribution obtained from the \textsc{MERCURIUS} FIR fitting for J154506 at $z=3.7515$ shown in Fig.~\ref{Fig:MERCURIUSfit} before correcting for magnification under a self-consistent scenario.
    Derived values for the dust mass (M$_{dust}$), opacity transition wavelength ($\lambda_{0}$), dust temperature (T$_{\rm{dust}}$), and dust emissivity index ($\beta_{\rm{IR}}$) are shown. Solid grey lines indicate the median (i.e. 50th percentile) of the parameter’s marginalised posterior distribution, while dashed lines show the 16th and 84th percentiles. In panels with the dust temperature, a dotted line indicates T$_{CMB}$ at the redshift of J154506 (T$_{\rm{CMB}}$=12.95~K). We note that the total IR luminosity (L$_{\rm{IR}}$) is not an independent parameter of the fitting routine, and is included purely for visualisation purposes. 
    }
\end{figure*}

\begin{figure*}[!h]
    \includegraphics[width=1.\textwidth]{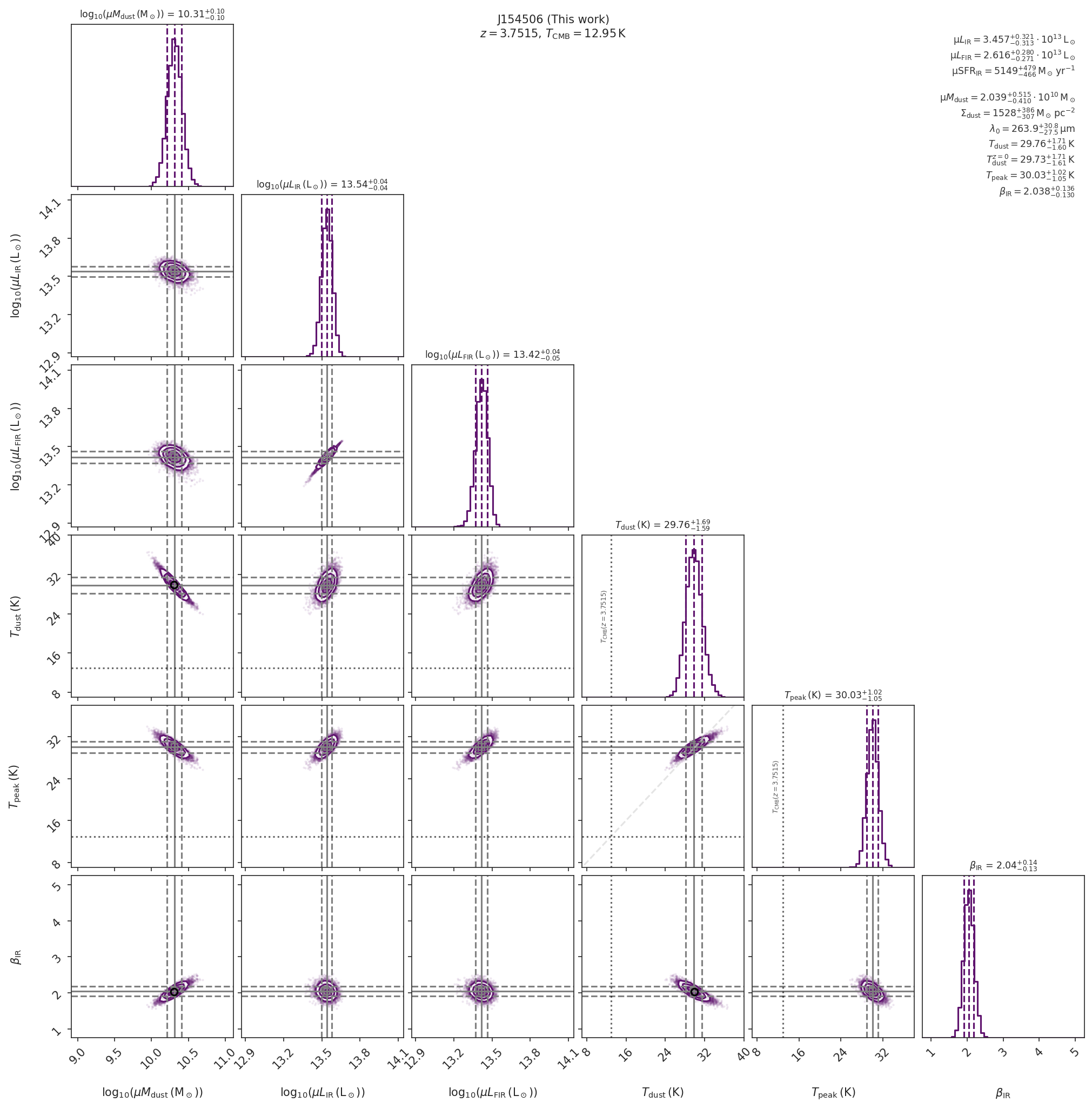}
    \caption{
    Corner plot of the posterior distribution obtained from the \textsc{MERCURIUS} FIR fitting for J154506 at $z=3.7515$ assuming an optically thin scenario.
    }
\end{figure*}

\clearpage

\end{appendix}

\renewcommand{\arraystretch}{1.3}
%-------------------------------------------------------------------
\end{document}